\documentclass[12pt]{iopart}
\usepackage{graphicx}
\usepackage{bm}
\begin{document}

\title[Transport properties in chaotic and non-chaotic many particles systems]
{Transport properties in chaotic and non-chaotic many particle systems}

\author{Fabio Cecconi$^1$, Massimo Cencini$^1$, and
  Angelo Vulpiani$^2$ }

\address{$^1$ INFM-CNR, Centre for Statistical Mechanics and
  Complexity (SMC) Dipartimento di Fisica, Universit\`a di Roma ``La
  Sapienza", Piazzale A.~Moro~2, 00185 Roma, Italy, and CNR-ISC, Via
  dei Taurini 19, 00185 Roma, Italy.}

\address{$^2$ Dipartimento di Fisica and INFN, Universit\`a di Roma ``La
  Sapienza", Piazzale A.~Moro~2, 00185 Roma, Italy}

\begin{abstract}
Two  deterministic models for Brownian motion are investigated by
numerical simulations and kinetic theory arguments. The first model
consists of a heavy hard disk immersed in a rarefied gas of smaller
and lighter hard disks acting as a thermal bath. The second is the
same except for the shape of the particles, which is now square.  The
basic difference in these two systems lies in the interaction:
hard-core elastic collisions make the dynamics of the disks chaotic
whereas that of squares is not.  Remarkably, this difference does not
reflect on the transport properties of the two systems: simulations
show that the diffusion coefficients, velocity correlations and
response functions of the heavy impurity are in agreement with kinetic
theory for both the chaotic and non-chaotic model. The relaxation to
equilibrium instead is very sensitive to the kind of interactions.
These observations are used to think back and discuss some issues
connected to chaos, statistical mechanics and diffusion.
\end{abstract}


\submitto{Journal of Statistical Mechanics: Theory and Experiments}
\maketitle

\section{Introduction}
\label{sec:intro}
A century after Einstein~\cite{Einstein} and Smoluchowski~\cite{Smolu}
seminal contributions, Brownian motion (BM) and diffusion phenomena
remain an active subject of research.  Statistical mechanics, since
its foundation, has been operating an elegant synthesis between the
microscopic dynamical laws and the macroscopic properties of a
system. In this context, Brownian motion is a paradigmatic example of
this \textit{modus operandi}.

In the statistical mechanics framework, the minimal condition needed
by microscopic dynamics for macroscopic diffusion can be identified in
the presence of a mechanism leading to velocity decorrelation --
memory loss.  In the effort of interpreting BM and non-equilibrium
transport in the light of modern dynamical systems theory, it thus
comes rather natural to identify in the chaotic character of
microscopic dynamics the main candidate to explain macroscopic
transport.  The instabilities of chaotic evolutions typically
produce irregular trajectories resembling Brownian motions and supply
a simple mechanism for memory loss. Indeed large scale diffusion has
been found in simple low dimensional chaotic
systems~\cite{DetDiff,DetDiff2,Klages}. This picture received
theoretical support from the existence, in some systems, of remarkable
quantitative relations between macroscopic transport coefficients --
such as diffusivity, thermal and electrical conductivity --, and
microscopic chaos indicators -- such as the Lyapunov exponents and the
Kolmogorov-Sinai entropy --, see e.g.
\cite{gaspnico,viscardy,GaspDor,dorfman,gaspard}.

On the other hand, non-chaotic models generating diffusion have been
proposed~\cite{ford2,dettmann,dettmann2,cecconi} raising some doubts
on the actual role of chaos for diffusion. However, these models,
representing possibly the most elementary examples of diffusion
without chaos, seem to be rather artificial: involve few degrees of
freedom, often the presence of quenched randomness is needed together
with the fine tuning of some parameters, e.g., for the suppression of
periodic orbits \cite{cecconi}. Their relevance to statistical
mechanics is thus not obvious. It is worth remarking that questions
about the relevance of chaos have been recently raised also in the
(related) context of thermal conduction problems~\cite{LLP}, where
non-chaotic models for heat transport were proposed and investigated
\cite{casati,casati2, bambihu,grassberger}.  

It should be remarked that the above systems are non-chaotic in the
sense that the Lyapunov spectrum is non-positive.  However, there are
well known examples of Lyapunov stable systems that display non
trivial
behaviours~\cite{ford2,cecconi,casati,casati2,bambihu,grassberger,fuzzy}.
In the presence of dynamical randomness without the sensitivity to
initial condition, as in quantum mechanics, an alternative definition
of "chaos" or "randomness" has been proposed in terms of the
positiveness of the Kolmogorov-Sinai entropy \cite{Gaspard1994}. In
classical systems with a finite number of degrees of freedom, as
consequence of the Pesin's formula, the two definitions
coincide. However, the proposal of Ref.~\cite{Gaspard1994} is an
interesting open possibility for quantum and classical systems in the
limit of infinite number of degrees of freedom.

It is far from trivial to the establish the role of chaos  
to macroscopic transport properties. The problem is
actually more general as chaos (in the above wider definition)
had been often invoked to justify the whole statistical mechanics
apparatus \cite{dorfman,gaspard} and to explain the irreversibility of
macroscopic processes \cite{prigogine}. Such viewpoint coexists with
the ``more traditional'' approach of Boltzmann, which stresses the
role of the many degrees of freedom and is mathematically supported by
the results of Khinchin~\cite{Khinchinbook}, Mazur and van der
Linden~\cite{Mazur} (see also Bricmont~\cite{Bricmont} and references
therein).

This paper aims to discuss the role of chaos in the context of diffusion,
thus we compare a chaotic and a non-chaotic many degrees of
freedom system, providing two distinct deterministic models for BM.  Both
models consist of rarefied gas of hard particles surrounding a larger
and heavier particle, in the following referred to as colloidal
particle (or colloid for brevity), impurity or test particle.  As an
effect of the large number of collisions, the deterministic motion of
such an impurity behaves as a BM at long times (i.e. much larger
than the collision time) provided its mass is much larger than that of
the lighter gas particles. The latter condition ensures a scale
separation between the gas and impurity dynamics. The asymptotic
convergence to a BM was proved for an infinite one dimensional system
of hard-core particles~\cite{Holley} and, later, in three dimensions
\cite{Lebovitz}.

\begin{figure}[t!]
\centering
\includegraphics[width=.4\textwidth,keepaspectratio,clip=true]{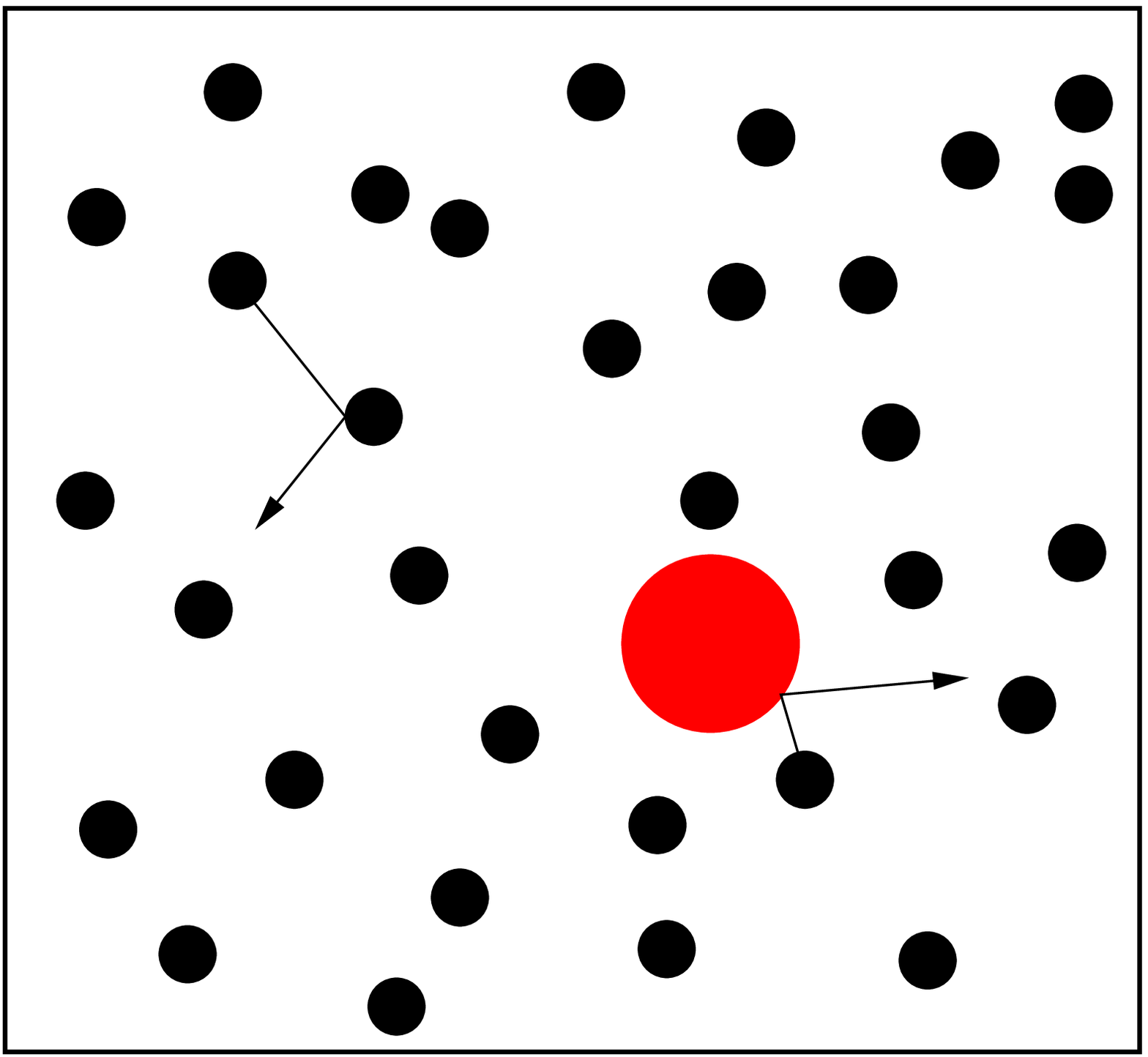}\qquad
\includegraphics[width=.4\textwidth,keepaspectratio,clip=true]{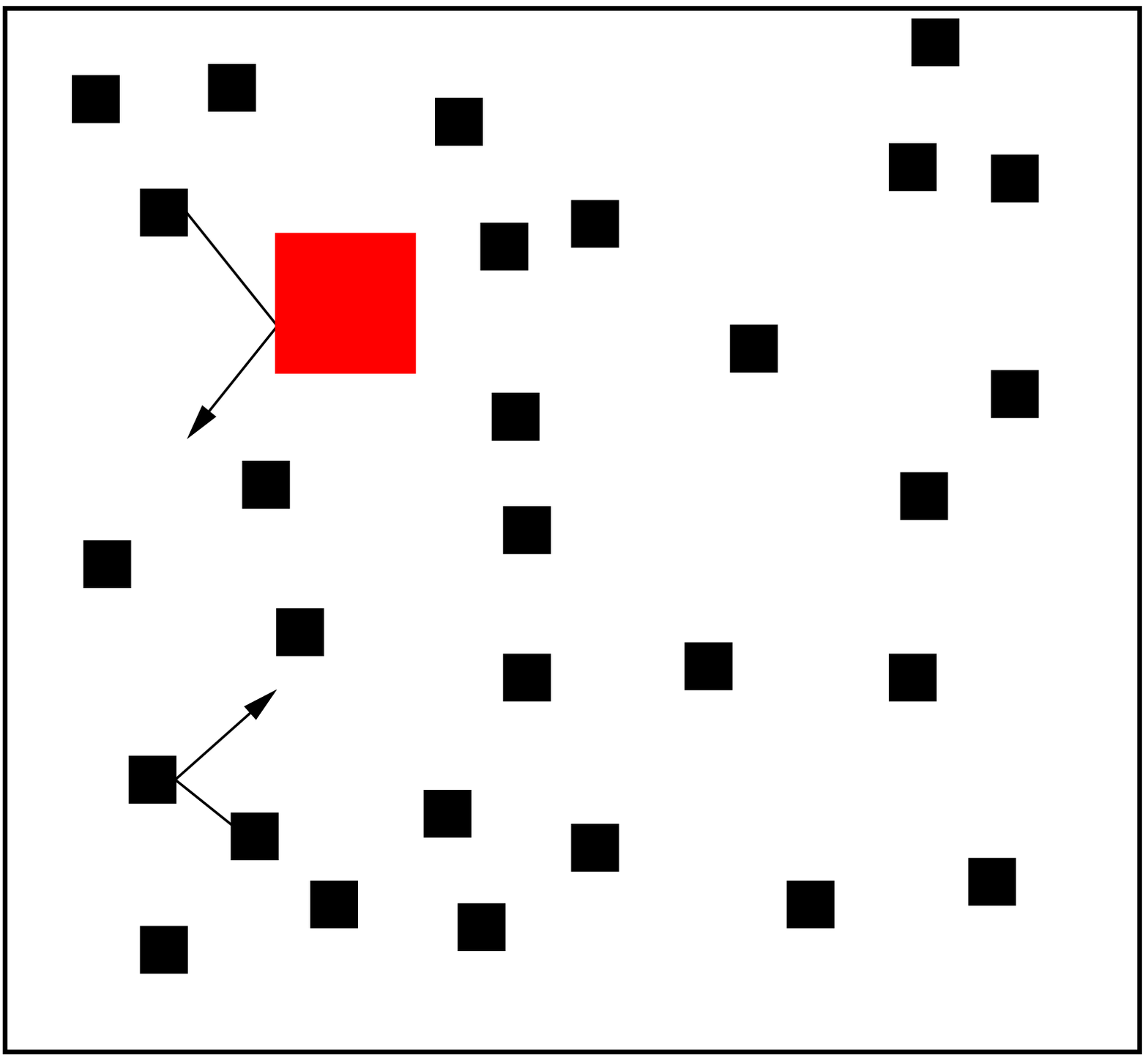}
\caption{Schematic illustration of two models used to discuss the 
mechanical diffusion: (left) hard disks and (right) hard parallel (non
  rotating) squares. The colloidal particle is displayed in red.}
\label{fig:illustration}
\end{figure}

The first model consists of $N$ hard disks surrounding a larger and
heavier disk as sketched in Fig.~1
(left). Particle interactions occur via binary elastic, hard-core
collisions which, due to convex shape of the disk, lead to chaotic
trajectories (i.e. exponential separation among initially close
trajectories).  In the following, this hard disk model will be denoted
by the acronym HD.

The second model, illustrated in Fig.~1 (right), consists of a gas of
$N$ hard (non-rotating) parallel squares (HPS) surrounding a larger and
heavier square particle.  In this case, binary hard-core elastic
collisions are not central and conserve the initial parallel
orientation of the squares. Unlike HD, the resulting dynamics is
non-chaotic.  This model is not new, the case of a HPS gas with
identical particles (without the impurity) has been studied both in
terms of molecular dynamics and kinetic theory by Frisch and coworkers
\cite{Frisch1,Frisch2,Frisch3,Frisch4}. It constitutes an interesting
statistical mechanics system characterized by the presence of an
infinite number of integrals of  motion, preventing the system from
being ergodic. Remarkably, in spite of such a pathology, the HPS gas
gives rise to, e.g., phase transitions and transport
properties as in the HD gas which, on the contrary, possesses
ergodic and mixing properties.

The colloid can be seen as a test particle to probe the transport
properties of both HPS and HD.  Notwithstanding the intrinsic
difference in their dynamics, chaotic for HD and non-chaotic for HPS,
their macroscopic diffusion properties are remarkably similar, as
confirmed by the analysis of the velocity-velocity correlations,
connected to diffusion via the Green-Kubo
formula~\cite{Kubo,kubobook}.  Measurements of
the response function of the velocity of the colloid and of gas
particles to small perturbations confirm that the non-chaotic nature
of HPS does not influence either the diffusion or the validity of the
Fluctuation Dissipation Theorem (FDT).

To discriminate the behavior of HPS from HD model one has to look at
the relaxation to equilibrium, when the initial state is very far from
it.  For square particles relaxation properties crucially depend on
the presence of the impurity, which induces a sort of ``effective
interaction'' among the gas particles and allows the system to
equilibrate to the Maxwell-Boltzmann distribution.  Since collisions
simply reshuffle the velocity components among the particles,
preserving the initial velocity distribution, without the colloidal
particle relaxation to Maxwell-Boltzmann is not possible. In spite of
such a constraint, self-diffusion and other statistical mechanics
behaviors can still be found \cite{Frisch1,Frisch2,Frisch3,Frisch4}
(see also Sect.~\ref{sec:diff}). In the presence of the colloid, a
signature of such a pathology survives: the relaxation time $\mathcal{T}_r$
strongly depends on the mass of the impurity, whilst in HD it is
independent of it. Moreover, ergodicity is not fully
recovered being the exchanges between the $x$ and $y$ components of
the velocity forbidden.

The paper is organized as follows. Sect.~\ref{sec:models} introduces
the two models.  In Sect.~\ref{sec:diff}, the numerical results for
the diffusion coefficient, self-diffusion and velocity
auto-correlation function are presented.  In Sect.~\ref{sec:relax},
the relaxation properties of HPS and HD close-to and far-from
equilibrium are analyzed. Sect.~\ref{sec:discuss} concludes the paper
with a discussion of the relation among chaos, statistical mechanics
and diffusion based on our results.  In the Appendix, a kinetic theory
derivation of the diffusion constants for both models is presented in
the very dilute limit.

\section{Deterministic many particle  models for diffusion}
\label{sec:models}
In the spirit of Smoluchovski's approach~\cite{Smolu}, it is natural to
introduce a mechanical model for BM based on the dynamics of a
macroscopically small but microscopically large heavy impurity
colliding with many lighter particles.  The large mass and size with
respect to the gas particles allows for a separation of time scales so
that the velocity of such an impurity is expected to follow a
Langevin equation.  The collisions with the gas provide both
the friction and the stochastic kicking to the
colloid~\cite{Grassia}. Here, we focus on two different models in two
dimensions.

\subsection{Hard disks model (HD)}
We consider $N$ hard disks of radius $r$ and mass $m$ plus an impurity
consisting in another disk of radius $R>r$ and mass $M>m$, as in
Fig.~1(left). All particles are in a square box of side $L$, with
periodic boundary conditions.  Crucial parameters are the number
density $\rho= N/L^2$ and the volume fraction $\psi= N\pi r^2/L^2$. In
the following, we shall always consider very dilute systems ($\psi\ll
1$) so that the properties of the system will be akin to those of a
rarefied gas.

The kinetic energy coincides with the total energy $H$
\begin{equation}
\label{II.1}
H= \sum_{j=1}^{N}   \frac{{\bm p}_j^2}{2m}  +\frac{{\bm P}^2}{2M} \equiv
\sum_{j=1}^{N+1}   \frac{{\bm p}_j^2}{2m_j}\,,
\end{equation}
 and is conserved.  In the second equality of Eq.~(\ref{II.1}) we
 adopt the convention that $i\!=\!N\!+\!1$ indicates the colloid,
 i.e. ${\bm p}_{N\!+\!1}\!=\!\bm P$ and $m_{N\!+\!1}\!=\!M$ while
 $m_i\!=\!m$ for $i=1,...,N$. Similarly, for the coordinates, we use
 either $\bm q_{i}=(x_i,y_i)$ ($i=1,\dots,N$) for the gas particles
 and $\bm Q=(X,Y)$ for the mass impurity, or ${\bm q}_j$ for
 $j=1,\dots N+1$ with ${\bm q}_{N+1}=\bm Q$. Of course, if $r=R$ and
 $m=M$, Eq.~(\ref{II.1}) reduces to a system of $N+1$ identical hard
 disks, which has been thoroughly investigated  by accurate
 computer simulations \cite{AT93} and theoretically in terms of
 kinetic theory of gases (see e.g. \cite{dorfman,REVIEWGENERALE} and
 references therein) in a variety of regimes.

Each particle moves with constant velocity till it collides with
another particle, event at which the velocities are updated according
to the elastic collisions rule
\begin{equation}
\label{eq:hdcrule}
{\bm p}_i' = {\bm p}_i + 
\frac{2m_i m_j}{m_i + m_j} ({\bm g}_{ij} \cdot \hat{\bm e}_{ij})\hat{\bm e}_{ij} \quad \textrm{and} \quad 
{\bm p}_j' = {\bm p}_j - 
\frac{2m_i m_j}{m_i + m_j}({\bm g}_{ij} \cdot \hat{\bm e}_{ij})\hat{\bm e}_{ij}
\end{equation}
where post-collision quantities are primed, and $\bm g_{ij} = \bm p_i/m_i 
-\bm p_j/m_j=\bm v_i-\bm v_j$ is the precollisional
relative velocity.  The unitary vector $\hat{\bm e}_{ij}$, oriented as
$i\to j$, connects the centers of the two disks at contact.

The model (\ref{II.1}) simplifies for two asymptotics: for $M/m\to 0$,
it approaches the Lorentz-gas model \cite{LorentzGas},
the impurity is much faster than the (now heavier) gas particles which
can be treated as immobile obstacles; for $R/r\to \infty$, the
Rayleigh-flight model is recovered if the $N$ small disks are very
dilute so that most of collisions involve the impurity.  In these two
limits, kinetic theory allows for analytical approximation of the
maximum Lyapunov exponent in terms of the system parameters
\cite{Posch}.

Here, being interested in Brownian motion, we consider situations in
which $R\gg r$ and $M \gg m$.  In such a limit, thanks to the large
time scale separation between the impurity and gas particles motions,
it is reasonable to assume  that each
velocity component $V$ of the colloid follows a Langevin
equation (we set $k_B=1$)
\begin{equation}
\label{II.4}
\frac{d V}{d t}= -\gamma V + \sqrt{2\gamma \frac{T}{M}} \, \eta\,,
\end{equation}
where $T$ is the gas temperature, and $\eta(t)$ a zero mean Gaussian
process with correlation $\langle \eta(t)\eta(t') \rangle
=\delta(t-t')$.  Of course, Eq.~(\ref{II.4}) describes the effective
dynamics of the impurity for times much larger than the average
collision time with the gas particles~\cite{Grassia,Holley,Lebovitz}.
The friction constant $\gamma$ sets the decay of velocity-velocity
correlation function
\begin{equation}
\label{II.8}
C_V(t)=\langle V(t)V(0) \rangle= \langle V^2 \rangle \mbox{e}^{-\gamma t}
=\frac{T}{M}\textrm{e}^{-\gamma t}\,,
\end{equation}
where brackets indicate time or ensemble averages.  By standard
kinetic-theory computation (see Eq.~(\ref{eq:gammaHD}) where also
corrections in $r/R$ and $m/M$ are taken into account) one has
\begin{equation}
\label{II.5}
\gamma= 2 \sqrt{2\pi}\, \frac{\rho R \sqrt{ mT}}{M}\,.
\end{equation}
Now, thanks to the Green-Kubo relation, linking the auto-correlation
function to the diffusion coefficient $D$
\begin{equation}
\label{II.7}
D=\int_0^{\infty} \langle V(t)V(0)\rangle dt = \frac{\langle V^2
  \rangle}{\gamma}=\frac{T}{M\gamma}\,, 
\end{equation}
it is straightforward to derive the diffusion constant of the
colloidal particle
\begin{equation}
\label{II.9}
D_c= \lim_{t\to\infty} \frac{1}{2t} \langle [X(t)-X(0)]^2 \rangle =
\frac{1}{2\sqrt{2\pi}} \, \frac{1}{\rho R}\sqrt{\frac{T}{m}}\,.
\end{equation}
Notice that $D_c$ is proportional to $\sqrt{T}$ and not to $T$ as in
liquids. This comes from the fact that in liquids the friction is
temperature independent, while in rarefied gases it is proportional to the
square root of the temperature.

Another interesting quantity, well defined both in the presence or
absence of the impurity, is the self-diffusion coefficient $D_g$ of a
tagged gas particle.  This is an important transport coefficient
associated with the gas density field evolution. Previous
investigations of hard disks and spheres
models studied in details such a quantity as well as other transport
coefficients (e.g., viscosity) at varying the parameters of the
problem.  In the dilute limit, the particle self-diffusion coefficient
takes the form \cite{dorfman}:
\begin{equation}
\label{II.3}
D_{g}= \lim_{t\to \infty} {1 \over 2t} \langle [x(t)-x(0)]^2 \rangle =
\frac{1}{4\sqrt{\pi}}\frac{1}{\rho r}\sqrt{\frac{T}{m}}\,,
\end{equation}
where $x$ indicates the $x$-component of the position of a tagged gas
particle.  At high volume fractions, corrections to the formula must
be considered.  Although the expression is formally similar to
(\ref{II.9}), only the prefactors change, the Langevin description
(\ref{II.4}) does not apply in this case due to the absence of time
scale separation. One has to take care to work in conditions in which
the coefficient (\ref{II.3}) is insensitive to the presence of the
impurity, so that it can be considered as a small perturbation.

Before passing to the other model, it is important to give a warning.
In two dimensions, as here, it is long known~\cite{Alder,Cohen_et_al}
that the gas velocity auto-correlation function develops small, slowly
decaying tails.  In principle, this may lead to an ill defined
diffusion coefficient. However, when $\psi \to 0$ this problem becomes
relevant only for enormously long times~\cite{longtimes}. Therefore, disregarding this
issue is justified from the practical point of view and we will
ignore it in the following.

\subsection{Hard parallel squares (HPS)}
We now consider a different model of hard-core particles in which the
$N$ gas disks are replaced by $N$ squares of side $2 r$ and mass $m$
plus a square colloid of side $2 R$ ($R>r$) and mass $M>m$, contained
in a box of size $L\times L$ with periodic boundary
conditions. 

 At the initial time, the sides of all squares are parallel (see
Fig~1(right)), and such parallelism is conserved
(no rotation) by the elastic, hard-core collisions, which are now not
central and amount to equal angle reflection against parallel
sides. The Hamiltonian (\ref{II.1}) is still describing the system,
and the collision rule (\ref{eq:hdcrule}) are replaced by
\begin{equation}
{\bm p}_i' = {\bm p}_i + \frac{2m_i m_j}{m_i + m_j}({\bm g}_{ij} \cdot \hat{\bm n}_{ij})\hat{\bm
n}_{ij} \quad \textrm{and} \quad {\bm p}_j' = {\bm p}_j - \frac{2m_i m_j}{m_i + m_j}({\bm
g}_{ij} \cdot \hat{\bm n}_{ij})\hat{\bm n}_{ij}\,,
\label{eq:collsquares}
\end{equation}
where ${\bm g}_{ij}$ is the relative velocity, and the unitary vector
$\hat{\bm n}_{ij}$ is directed as the normal to the colliding sides,
that is, either along the $x$ or $y$ axes.   At variance with the
HD, the HPS model is not a mechanical system as its evolution does not
follow the Newtonian dynamics, because of the constraint keeping the
parallelism among the square sides.  Since the collision rules
(\ref{eq:collsquares}) are linear, the evolution law for the tangent
vector coincides with that of the system and the Lyapunov exponents
are all equal to zero. Therefore, unlike the case of disks, the system
is non-chaotic, although the presence of the corners may induce
non-linear instabilities, producing a defocusing for non-infinitesimal
displacement among two trajectories.

In the absence of the impurity the considered system reduces to $N$
identical parallel squares, which was studied by Frisch and coworkers
\cite{Frisch1,Frisch2,Frisch3,Frisch4} for its peculiar
properties. This system is indeed interesting under several aspects.
First of all, due the rules of interaction (\ref{eq:collsquares}),
colliding particle pairs simply exchange their velocities along the
direction of impact: similarly to $1$d hard rods a collision event
corresponds to a relabeling of particles, though in the $2d$ case as
here the relabeling is only for one component of the velocity.  This
implies that all velocity moments are conserved separately for the two
components. It thus follows that a velocity probability distribution
function (pdf), which is initially factorized
$P(v_x,v_y;t=0)=p(v_x)p(v_y)$, is preserved by the time evolution.  In
other words, unlike HD, the system is non-ergodic and the pdf of the
velocities cannot relax to the Maxwell-Boltzmann
distribution. However, if the initial distribution is not factorized,
it will factorize $P(v_x,v_y;t\to \infty)=f(v_x)g(v_y)$ with $g(v_x)=\int dv_y
P(v_x,v_y;t=0)$ and $g(v_y)=\int dv_x P(v_x,v_y;t=0)$. Indeed the relabeling
induced by the collisions decorrelate the $x$ and $y$ components.

In spite of these pathologies, simulations~\cite{Frisch1} and kinetic
theory computations~\cite{Frisch2} show that this non-ergodic system
possesses many properties akin to those of the HD, which instead is
typically considered to be ergodic and mixing.  For instance,
transport properties are well defined and in the dilute limit.  The
self-diffusion coefficient for an initial Maxwell-Boltzmann
distribution can be computed~\cite{Frisch1}
\begin{equation}
\label{II.10}
D_g= \frac{\sqrt{\pi}}{16} \frac{1}{\rho r}\sqrt{\frac{T}{m}}  \, .
\end{equation}
Moreover, phase transitions similar to those found in HD can be
observed for suitable values of the volume fraction $\psi=
N4r^2/L^2$ \cite{Frisch2}.

When the impurity is present the situation changes. Due to the mass
difference ($m\neq M$), many conserved quantities are destroyed and
collisions can change the velocities pdf: relaxation to
Maxwell-Boltzmann distribution is now possible.  Essentially the
impurity allows for an ``effective interaction'' as
e.g. in~\cite{LeboCherno}. However, ergodicity is not fully recovered:
it is clear that, due to the collision rule, no mixing between the
velocities along $x$ and $y$ is possible. Meaning that
$E_x=\sum_{i=1}^{N+1} {p}_{xi}^2/2m_i$ and $E_y=\sum_{i=1}^{N+1}
{p}_{yi}^2/2m_i$ are separately conserved.  We shall always work in
isotropic conditions, i.e. with initial velocity pdf such that
$E_x=E_y$

In such a condition, being an equilibrium distribution well defined,
we can still consider the limits $R \gg r$ and $M \gg m$ and study the
BM of the impurity. From kinetic theory (see Eq.~(\ref{eq:dcHS}) in
the Appendix), one can compute the friction coefficient
\begin{equation}
\label{gammaHSpaper}
\gamma= 8\sqrt{\frac{2}{\pi}} \frac{\sqrt{m}\rho R \sqrt{T}}{M}\,,
\end{equation}
and therefore the diffusion constant
\begin{equation}
\label{II.12}
D_c= \frac{\sqrt{\pi}}{8\sqrt{2}} \frac{1}{\rho R}\sqrt{\frac{T}{m}}  \,\, .
\end{equation}
Comparing the above expression with (\ref{II.9}) reveals that the only
difference lies in a numerical prefactor which takes into account the
different geometries of the particles.

\section{Transport Properties}
\label{sec:diff}
Let us now investigate, by means of numerical simulations, the
transport properties (diffusion of the impurity and self-diffusion of
tagged gas particles) in HD and HPS models both in the presence and
absence of the impurity. 

Before discussing the results, we briefly summarize the numerical
methods employed. For both HPS and HD, we used an event driven
algorithm with the minimum image convention~\cite{AT93}. All
simulations refer to the rarefied case. After several tests we fixed
$\rho =N/L^2=10$ and solid fraction $\psi\approx
\mathcal{O}(10^{-3})$, so that we can neglect all known corrections to
the transport coefficients \cite{REVIEWGENERALE}. The diffusion
constants are then measured in terms of the mean square displacement
in the $x$ and $y$ direction for both the colloid and tagged gas
particles, see Eqs.~(\ref{II.9}) and (\ref{II.3}).  Working in
isotropic conditions, the horizontal and vertical diffusion
coefficient are equal, allowing us to average over the two components
for increasing statistics.  Averages have been computed over
$\mathcal{N}$ time windows in a long run. The diffusion constant $D_c$
is thus obtained by looking at
\begin{equation}
\langle [\Delta Q(t)]^2 \rangle = \frac{1}{\mathcal{N}}\sum_{k=1}^{\mathcal{N}}
[Q(t+t_k)-Q(t_k)]^2 \;\approx 2D_c t\,,
\label{eq:incre}
\end{equation}
where $Q=X,Y$ represents any of the components of the particle
position, and $0\leq t \leq (t_{k+1}-t_k)$. The length of the window
$t_{k+1}-t_{k}$ is chosen long enough to observe a diffusive regime
over about two decades.  For the self-diffusion we proceeded
similarly, but working on the gas particle positions,
$q_i=x_i,y_i$. Being all gas particles equivalent, we average over
them
\begin{equation}
\langle [\Delta q(t)]^2 \rangle \!=\!
\frac{1}{N\mathcal{N}}\sum_{k=1}^{\mathcal{N}} \sum_{i=1}^{N}
     [q_i(t+t_k)-q_i(t_k)]^2 \;  \approx 2D_gt\,
\label{eq:selfdiff}
\end{equation}
with abuse of notation the same brackets as in (\ref{eq:incre}) are
used though in (\ref{eq:selfdiff}) the average over all particles is
also performed.
HD and HPS simulations differ only for the collision rules.  While
computing the average displacements we also computed the velocity
auto-correlation functions for both the colloid and gas particles
$C_V(t)$ and $C_v(t)$. The averages have been performed, as for the
displacements, over different time windows and, for the gas, also over
all particles.

We work with a single impurity with $R=10r$ and $M=20m$ (tests with
different sizes and masses have been performed). Due to the necessity
of averaging over many, typically $\mathcal{N}=10^3$, time windows we
employ a not too large number of gas particles ($N=300$, tests with
$N=1000$ have been done).  Using this relatively low number of
particles and the mass ratio $M/m=20$ requires to take into account
finite size corrections\footnote{In practice, the temperature of the
  impurity will be smaller by a factor $1-M/M_T$ than that of the gas
  (with $M_T=\sum_{i=1,N+1}m_i$). Other corrections by terms $m/M$ and $r/R$ 
should be included as well, see Appendix.}.  Indeed the constraints of energy
and momentum conservation (the latter ensured by the periodic boundary
conditions) are known to cause a breakdown of the energy equipartition
for simulations with non-equal mass particles \cite{A93,SBJ06}.

\begin{figure}[t!]
\includegraphics[width=.48\textwidth]{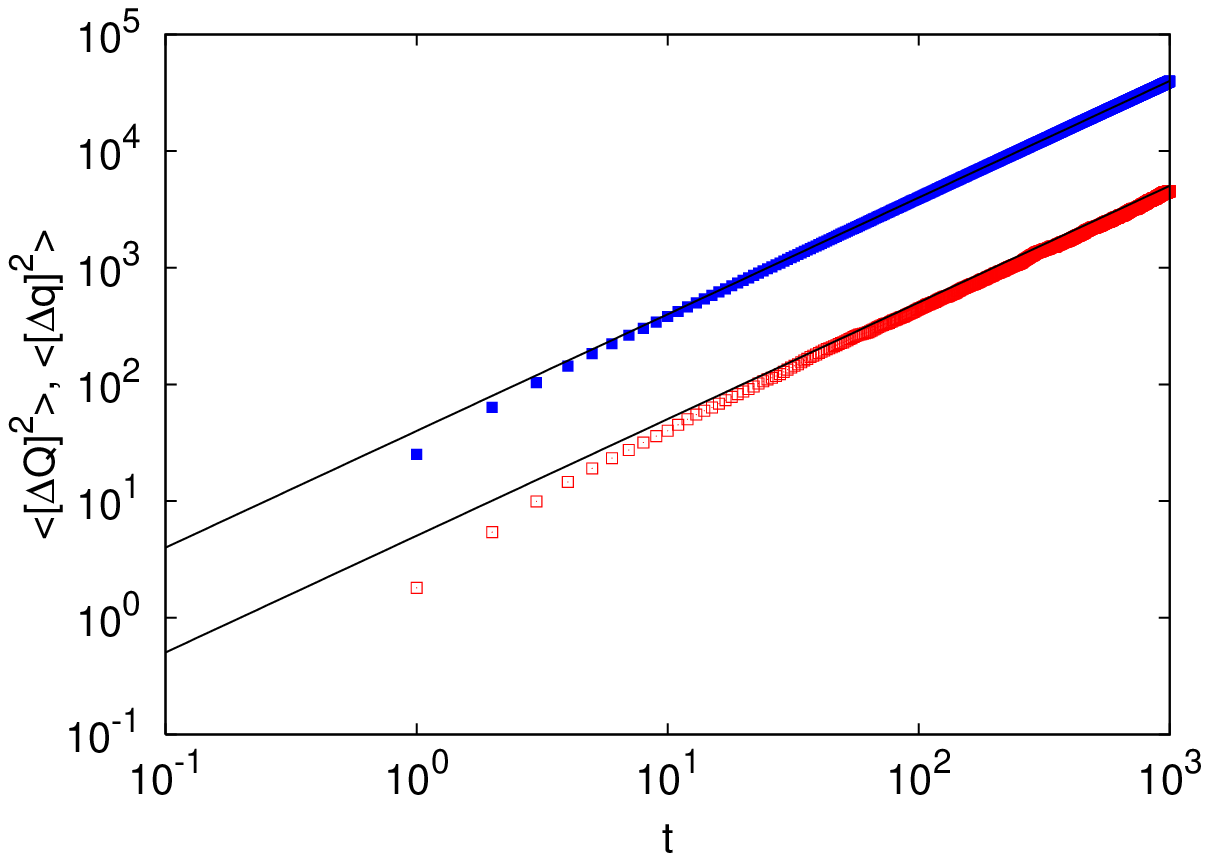}\hfill \includegraphics[width=.48\textwidth]{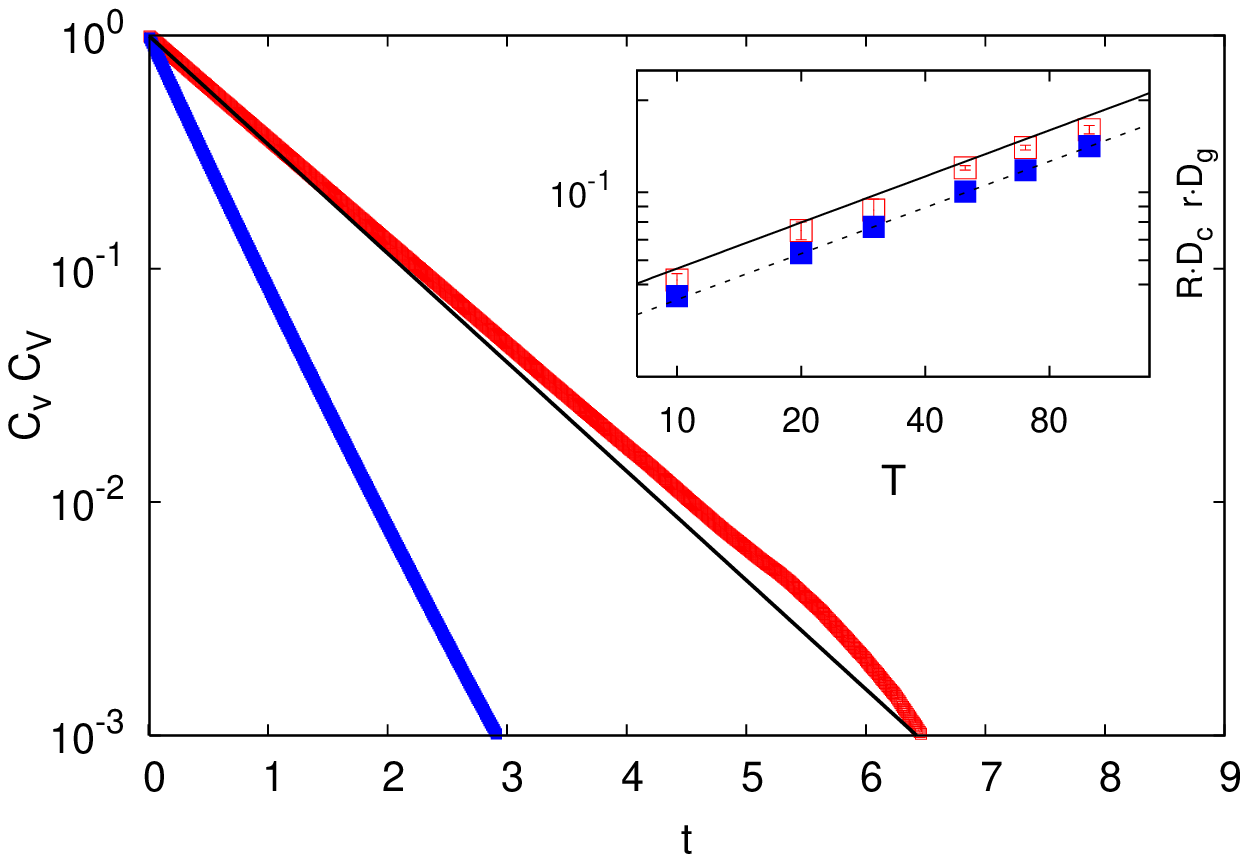}
\centerline{\qquad (a)\hspace{7.5cm}(b)}

\caption{(a) Mean square displacement for the colloid $\langle [\Delta
Q(t)]^2\rangle$ (red) and tagged gas particles $\langle [\Delta
q(t)]^2\rangle$ (blue) vs time for HD. The straight lines show the
expectations (\ref{II.9}) and (\ref{II.3}), respectively. The
temperature is set to $T=50$.  (b) Velocity auto-correlation for
colloid $C_V(t)$ (red) and gas particles $C_v(t)$ (blue) measured in
the same simulation as (a).  The straight line has a slope given by
the friction constant (\ref{II.5}).  Inset: Normalized diffusion
constants $RD_c$ and $rD_g$ vs $T$. Solid and dotted lines correspond
to the expectation (\ref{II.9}) and (\ref{II.3}) with the finite
sample size corrections included~\protect\cite{A93,SBJ06}.  $N=300$
gas particles of radius $r=0.005$ and mass $m=1$ have been used, with
number density $\rho=10$, for the impurity $R=10r$ and
$M=20m$. Averages are performed over $\mathcal{N}=1000$ time
windows. Simulations on a shorter time scale with $N=1000$ particles
give the same result within statistical errors.  Notice that the self
diffusion is by definition the results of a further average over the
number of gas particles, thus its statistics is much improved than
$D_c$ implying lower statistical errors. }\label{fig:diffdisks}
\end{figure}

We are now ready to discuss the results. Let us start from
Fig.~\ref{fig:diffdisks}a where we show the data for the diffusion and
velocity correlation of the impurity and tagged gas particles in the
case of hard disks.  Although the diffusive properties of HD have
been thoroughly studied in the literature, they serve for comparison
with  HPS, presented below.  As one can see, the diffusive
scaling extends over about two decades allowing for a good estimate of
the diffusion coefficients, which are in agreement with the
expectation (\ref{II.9}) and (\ref{II.3}), see also the inset in
Fig.~\ref{fig:diffdisks}b.  The auto-correlation functions $C_{V}$ and
$C_{v}$ (shown in Fig.~\ref{fig:diffdisks}b) display an exponential
decay. For the colloid the decay time is in agreement with the
friction constant (\ref{II.5}), making the approximation of the
colloid evolution in terms of a Langevin equation meaningful.

\begin{figure}[t!]
\includegraphics[width=.48\textwidth]{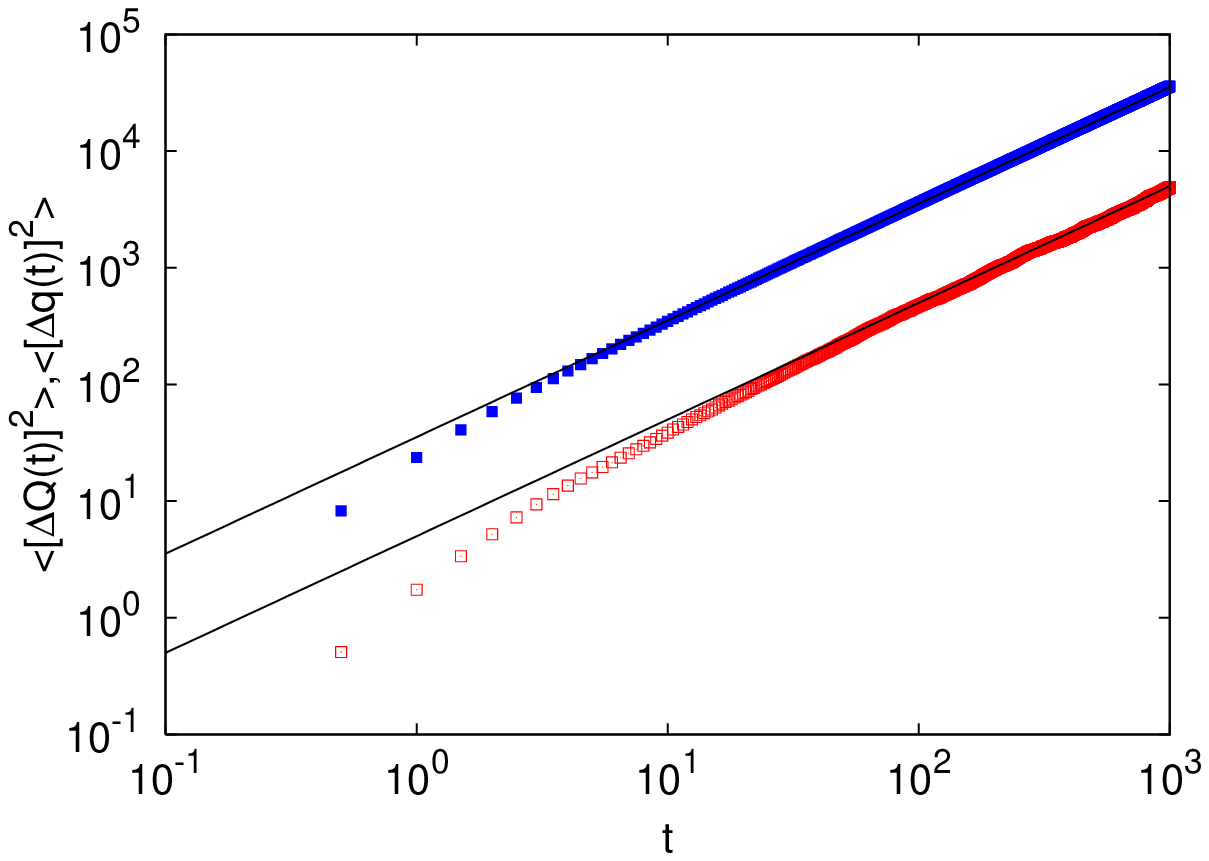}\hfill 
\includegraphics[width=.48\textwidth]{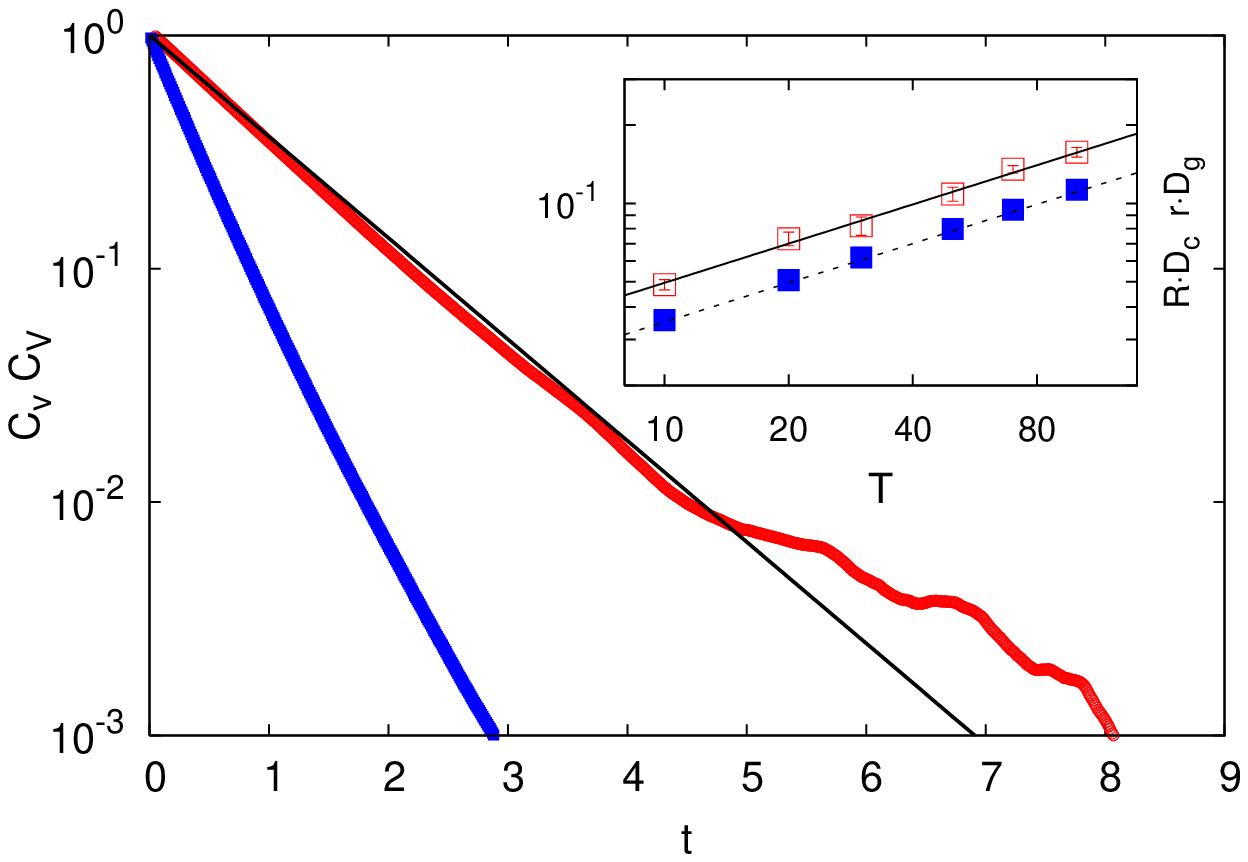}
\centerline{\qquad (a)\hspace{7.5cm}(b)}
\caption{Same of Fig.~\protect\ref{fig:diffdisks} for HPS. The size of
  the particles has been chosen so to have the same area of the
  disks.}
\label{fig:diffsquares}
\end{figure}

We now consider the case of $N$ parallel squares with a mass impurity.
As discussed in the previous section, the presence of the colloid
allows for the existence of an equilibrium state characterized by the
Maxwell-Boltzmann distribution for the particle velocities. However,
unlike HD, the system is non-chaotic. Moreover, ergodicity is broken and there is no
mixing between the horizontal and vertical components.  As
Fig.~\ref{fig:diffsquares}a clearly shows the transport properties of
HPS are well defined, the colloidal particle and tagged gas particles
diffuse with coefficients as given by Eq.~(\ref{II.12}) and
(\ref{II.10}), respectively; see also inset in
Fig.~\ref{fig:diffsquares}b. Moreover, the system loses memory as both
the velocity auto-correlation function of the impurity and gas
particles decay exponentially (Fig.~\ref{fig:diffsquares}b). At a
first sight, the agreement of the self-diffusion constant with
(\ref{II.10}) may appear strange. Indeed, the formula derived by
Frisch and collaborators~\cite{Frisch1} refers to a gas of identical
particles, where no equilibration occurs. However, Eq.~(\ref{II.12})
was obtained assuming a Gaussian distribution for the velocities, here
always realized thanks to the presence of the impurity which allows
for Maxwell-Boltzmann distribution.

We consider now the model in which no impurity is present, previously
studied by Frisch and collaborators \cite{Frisch2,Frisch3}. As
stressed above, in such a case HD and HPS (i.e. chaotic and
non-chaotic particle systems) are conceptually very different: while
HD remains a well defined statistical mechanics system with relaxation
to an equilibrium state independent from the initial conditions, HPS
never reaches equilibrium, the initial velocity distribution remains
unchanged with time. Nevertheless both systems display well defined
transport properties.

\begin{figure}[t!]
\centering
\includegraphics[width=.6\textwidth]{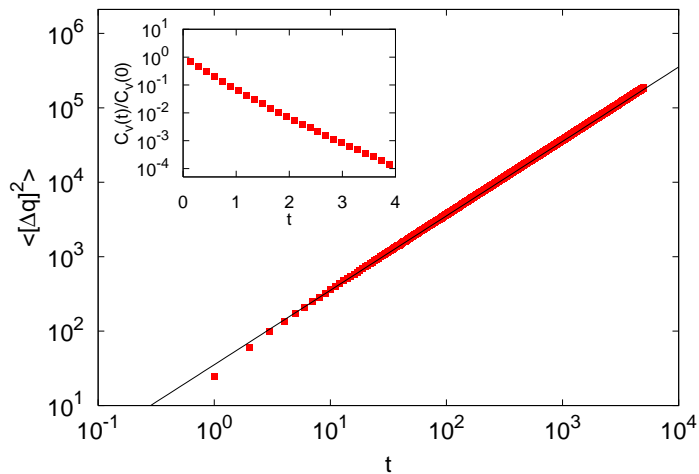}
\caption{(a) $\langle \Delta q^2(t)\rangle$ vs $t$ for a gas of
  $N=300$ hard parallel squares with size as in
  Fig.~\protect\ref{fig:diffsquares} and with a Gaussian velocity
  distribution at temperature $T=50$ and number density
  $\rho=10$. Inset: velocity auto-correlation function. With $N=1000$
  we obtained indistinguishable results.}\label{fig:selfsquare}
\end{figure}

 By computing $D_g$ for HD in the same conditions as
Fig.~\ref{fig:diffdisks}a we checked whether the value of $D_g$ agrees
with its measurement done in the presence of the colloid. We actually
found a perfect agreement, within errors, between the two measurements
(not shown).  This confirms \textit{a posteriori} that the colloid can
be considered a small perturbation for the gas without important
consequences for the transport properties of the HD gas. More
interesting is to investigate the case of squares.

In Fig.~\ref{fig:selfsquare}, we show $D_g$ for a gas of equal hard
squares with Gaussian initial distribution having the same temperature
as in the simulations with the impurity: the diffusive behavior is
again a robust well defined property. It is worth underlining the
following points.  First, a diffusive behavior is observed also
also in the absence of relaxation to a statistically steady
state. Second, the value of $D_g$ matches the theoretical value
obtained by Frisch \cite{Frisch1} and is in perfect agreement, within
errorbars, with the equivalent quantity obtained in the presence of
the impurity. Third, the velocity-velocity correlation function, shown
in the inset, decorrelates exponentially with a good degree of
approximation.

The presence of diffusive behaviour in non-chaotic systems is not new.
Models consisting of non interacting particles has been already
considered in Refs.~\cite{dettmann,dettmann2}. However, such models,
though interesting from a dynamical system point of view, consist of
independent particles and thus are somehow far from a statistical
mechanics perspective.  The HPS system here investigated is
non-chaotic but having many degrees of freedom in interaction it
constitutes a valuable statistical mechanical system.

We conclude this section by stressing that, although the presence of
the colloid changes conceptually the properties of the system,
transport properties in the presence or absence of the impurity are
quantitatively the same, provided the Gaussian distribution is
chosen. This means that, at least, for transport properties the
colloid can be considered as a small perturbation to the system,
but for a non generic state.

\section{Relaxation Properties} \label{sec:relax}

The analysis of the relaxation processes associated with spontaneous or
induced statistical fluctuations represents the classical approach to
probe the macrostates explored by a system. For instance, Fluctuation
Dissipation Theorems (FDT) \cite{kubobook}, relating the behavior of
spontaneous fluctuations at equilibrium to the average response of a
system to infinitesimal perturbations, establish a connection between
equilibrium (correlation functions) and non-equilibrium (response
functions) quantities.  We carried out a set of simulations to
determine the relaxation properties of the system both close-to the
equilibrium state, characterized by a Maxwell-Boltzmann distribution,
and far-from it (e.g., starting from a uniform distribution) to
probe the possible influence of chaos on the statistical properties.
  
\subsection{Close to equilibrium}
To gain information about transport properties by studying the
relaxation to equilibrium it is useful to introduce the response
function to small impulsive perturbations applied to a component $V$
of the velocity of the impurity. This can be obtained, e.g., by
applying a force $f(t)=F\delta(t)$ which acts only at $t=0$ with $F\ll
1$. The result of $f(t)$ is to cause an instantaneous (very small)
variation of the velocity $V(0)\to V(0)+\delta V_0$, with $\delta
V_0=F/M$. One can thus define the average response function as $R_V(t)
= \langle \delta V(t)\rangle_e/\delta V_0$, where $\langle [\dots]
\rangle_e$ denotes an ensemble average at fixed time in the presence
of an impulsive perturbation. The response $R_V(t)$ bears important
information on the transport properties of the system being.
Indeed,  if the velocity distribution is Gaussian the classical FDT
relation \cite{kubobook}  tells us that $R_V(t)$ coincides with the
normalized velocity correlation function
\begin{equation}
\label{eq:fdt3}
R_V(t) = \frac{C_V(t)}{C_V(0)}\,.
\end{equation}
This can be seen as the differential form of the Einstein relation
connecting the asymptotic speed of the particle to the mobility under
the effect of an infinitesimally small force.
In the following, we present the measurement of the response
function $R_V(t)$, which will be compared with the correlation
$C_V(t)$ to test whether FDT holds. 
Similarly, we analyze also the response function of a
tagged gas particle $R_V(t)$ when the impurity is absent.

In order to numerically compute the response function, we adopted the
following protocol.  Consider a system HPS or HD in the presence of
the colloidal particle and let it evolve till the equilibrium state is
reached. At this time, that we call $t=0$, the velocity of the
colloidal particle is perturbed by a small amount $V(0) \to V(0) +
\delta V_0$. In principle, the perturbation should be infinitesimal
(i.e. $\delta V_0\to 0$). This is however infeasible in practical
computations, we then considered three different perturbation values
defined in fraction of the root mean square velocity $\delta V_0 =
\alpha \sqrt{T/M}$ with $\alpha=0.02,0.05$ and $\alpha=0.1$. In this
way, comparing the different numerical experiments, we can test {\it a
posteriori} the validity of the linear response theory.  This small
perturbation on the velocity is expected to be re-adsorbed, meaning
that the velocity of the impurity should, after a while, assume values
drawn from the equilibrium distribution.  The procedure is thus
repeated for many (typically $10^{4}-10^{5}$) times. The time history
of the colloidal particle $V(t)$ is followed in each experiment so to
obtain its average evolution $\langle V(t)\rangle_e $, from which the
response function $R_V(t)$ can be defined as
\begin{equation}
R_V(t) = \frac{ \langle V(t) \rangle_e}{\delta V_0}\,,
\label{eq:fdt}
\end{equation}
note that we used  that $\langle V(0)\rangle_e =0$. 

\begin{figure}[t!]
\centering
\includegraphics[width=0.5\textwidth]{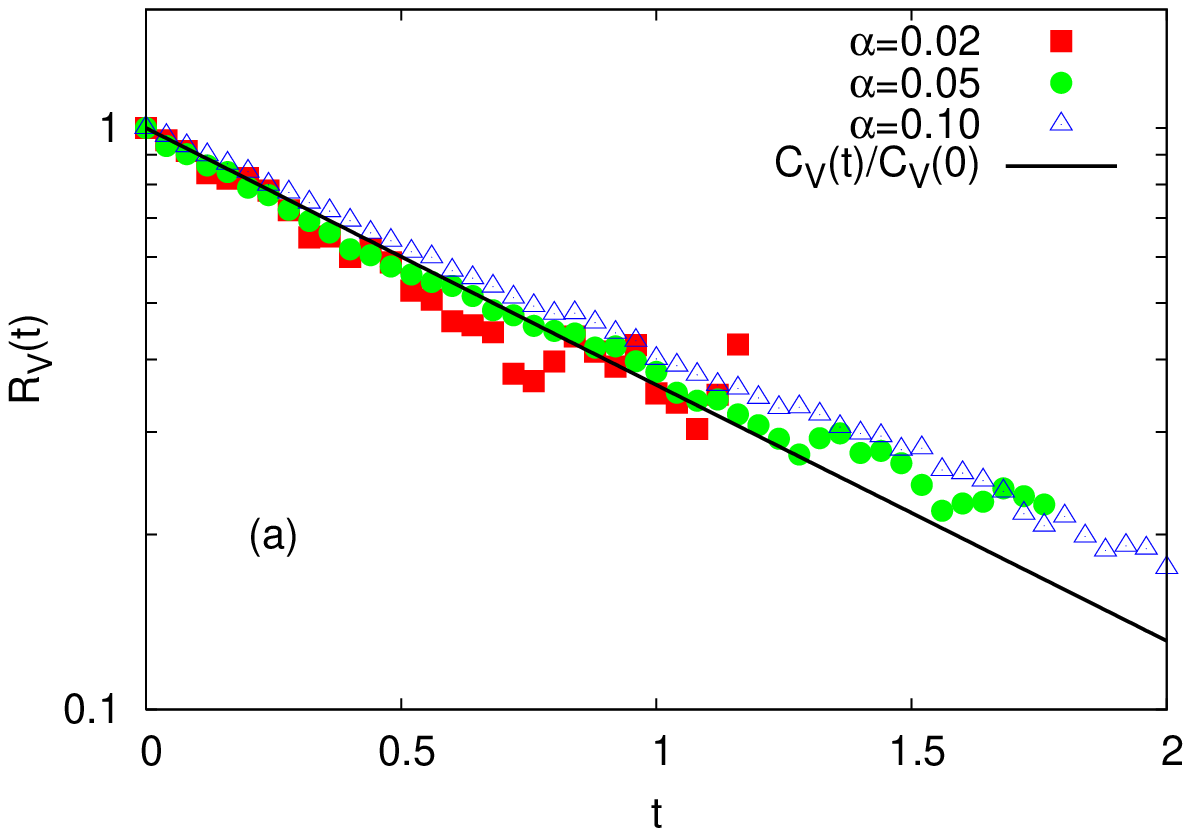}\hfill\includegraphics[width=.5\textwidth]{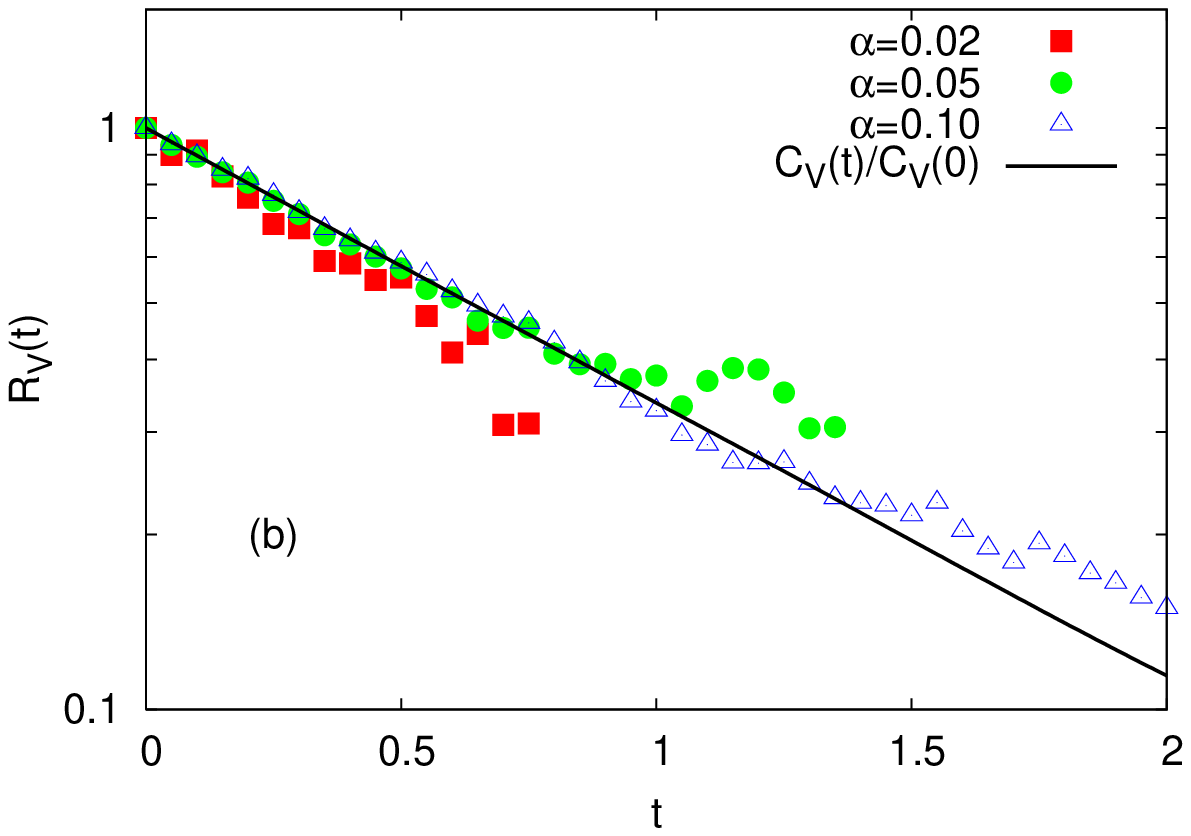}
\caption{(a) $R_V(t)$ vs $t$ measured for $\delta
  V_0=\alpha\sqrt{T/M}$ with $\alpha=0.02,0.05,0.1$ as in the legend
  and $C_V(t)/C_V(0)$ (solid line) computed in the same conditions.
  The parameters are as in Fig.~\protect\ref{fig:diffdisks} with
  $T=50$.  Note the good collapse, within statistical errors, of the
  response functions for $\alpha=0.02$ and $\alpha=0.05$ and the
  correlation function. Deviations can be appreciated for $\alpha=0.1$
  (presumably for such a value of the perturbation one exits the
  linear response regime).  (b) the same of (a) for HPS system with
  parameters as in Fig.~\protect\ref{fig:diffsquares}, the agreement
  between the decay of the correlation and the response
  function is also in this case very good. For both (a) and (b), the
  three curves have different length because decreasing the initial
  perturbation value the signal is spoiled by noise at earlier times.
}
  \label{fig:fdtsquare}
\end{figure}

Equation (\ref{eq:fdt}) has been verified in our simulations for
both HPS and HD, and the comparison between the numerical results is
shown in Fig. \ref{fig:fdtsquare}. One should notice first that for
both HD and HPS the response measured for different perturbations
superimpose, meaning that we are in the linear response
regime. Moreover, the fair superposition of the exponential decays of
$C_{V}(t)$ and $R_V(t)$, up to statistical errors, constitutes the numerical
evidence for both systems to obey FDT relation.

The conclusion that can be drawn is that whenever the chaotic HD and
the non-chaotic HPS are prepared into an equilibrium state compatible
with the thermodynamic parameters $T$ and $\rho$, the behavior of
$R_V(t)$ does not reveal any difference between the two systems.

\begin{figure}[t!]
\centering
\includegraphics[width=0.5\textwidth]{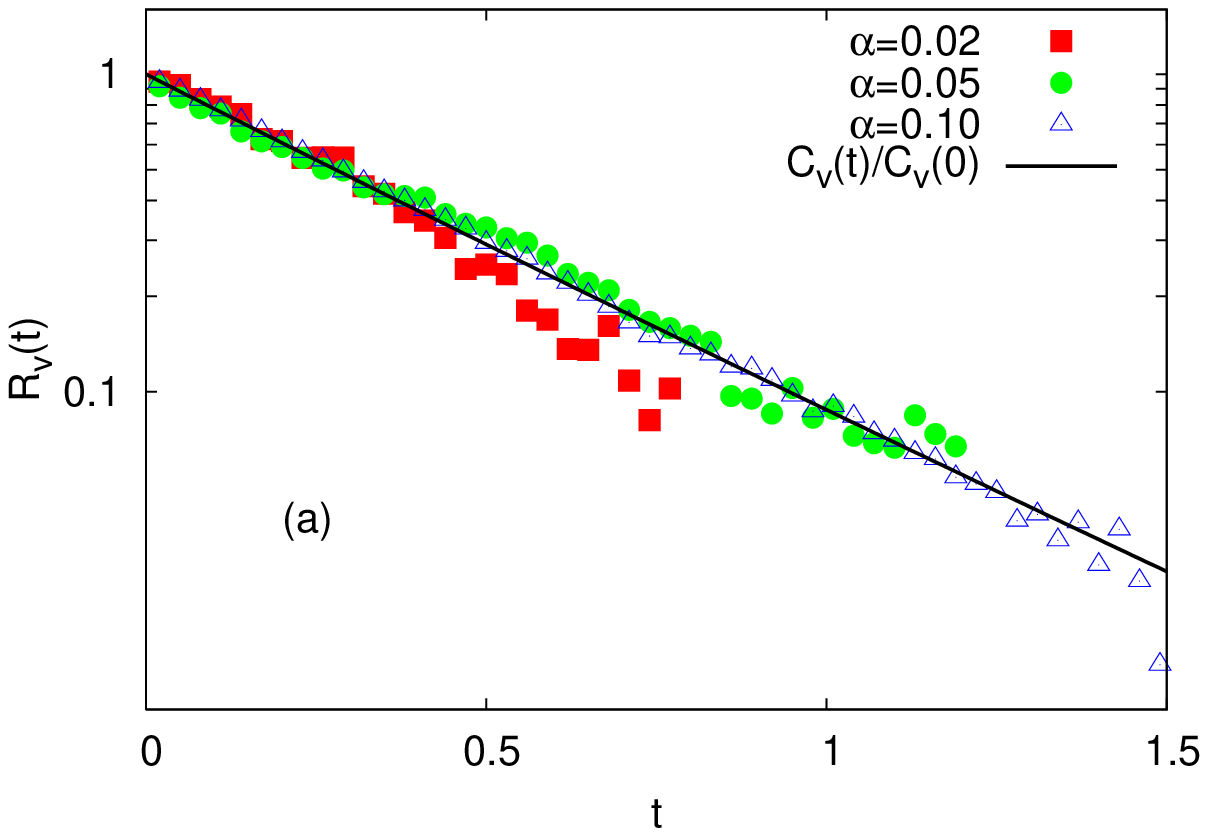}\hfill
\includegraphics[width=0.5\textwidth]{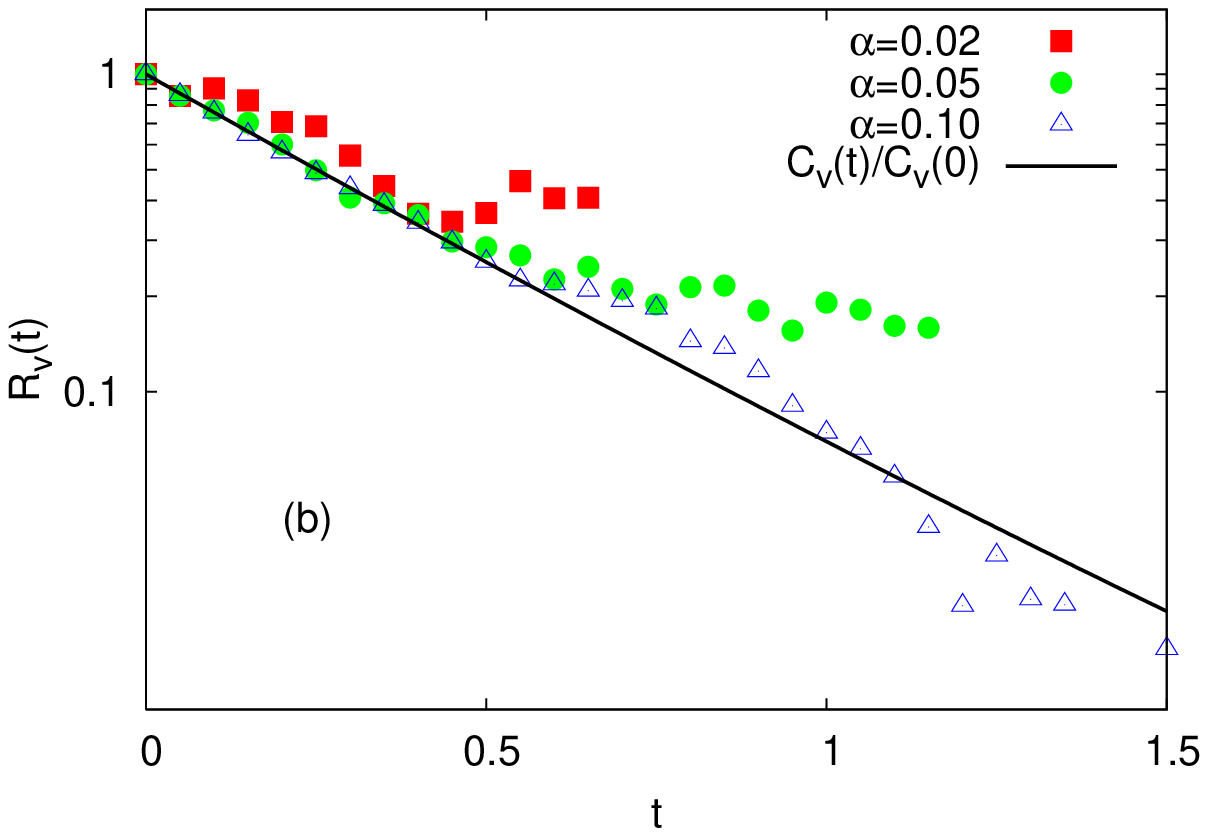}
\caption{The same of Fig.~\protect\ref{fig:fdtsquare}(a) showing the
FDT relation for the HD (a) and HPS (b) gas particle (in the absence of
the impurity).  For HPS the initial distribution of the gas velocities
was chosen Gaussian with temperature $T=50$ so to compare with the
correlation function as measured in
Fig.~\protect\ref{fig:selfsquare}.}
  \label{fig:fdt2}
\end{figure}

The above procedure can be applied to a tagged particle so to define
the response for gas particles $R_v(t)$ which is connected to the
correlation function $C_v(t)$. We perform such a measurement for the
system of HD and HPS without the impurity.  Being the perturbation
very small the measurement is still meaningful though the
system is not equilibrating.

Figure \ref{fig:fdt2} shows the average response functions for a
tagged particle in both HD and HPS, together with the comparison with
the correlation functions. There is a fair agreement between $R_v(t)$
and $C_v(t)/C_v(0)$ for both HD and HPS models. Of course, in the case
of the HPS the origin of the validity of a FDT relation cannot be
ascribed to the presence of chaos. Clearly it can only come from the
presence of many degrees of freedom and should have a probabilistic
origin.

\subsection{Far-from equilibrium}
\label{sec:relaxfar}
We consider now the case of relaxation when the system is prepared
into a state far-form equilibrium, for which the difference between
the two models becomes evident.  This can be understood from the
outset by recalling that HPS systems with identical squares cannot
relax due to the collision rules that merely relabel the velocity
components.  Unlike HPS, hard disks collisions, also thanks to chaos,
mix the velocity components at each impact and allow for a fast
relaxation to a Maxwell-Boltzmann distribution.

When the colloid is introduced, relaxation to equilibrium becomes
possible also in HPS because impacts against the impurity break the
relabeling process and provide a mechanism for the transfer of energy
(at least, separately for the $x$ or $y$ components which do not mix),
this is similar to the problem of the adiabatic
piston considered in Ref.~\cite{LeboCherno}.  However, since only collisions with the
impurity contribute to the relaxation process of the gas, the
time-scale for reaching the equilibrium state crucially depends on its
mass $M$ and size $R$.  This contrasts with the HD model, for which
the time scale of relaxation is essentially unaffected by the
characteristics and/or the presence of the colloid.
   
\begin{figure}[t!]
\includegraphics[width=.5\textwidth]{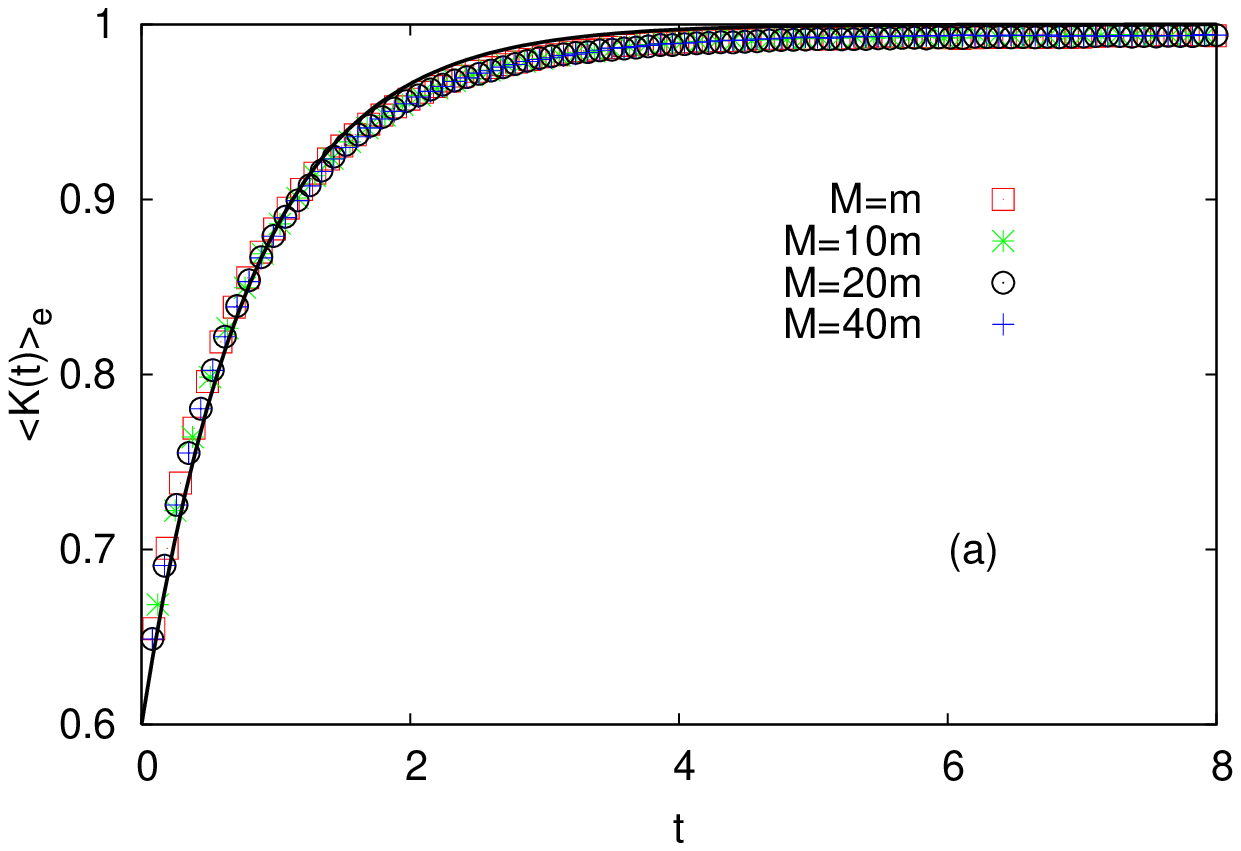}\hfill
\includegraphics[width=.5\textwidth]{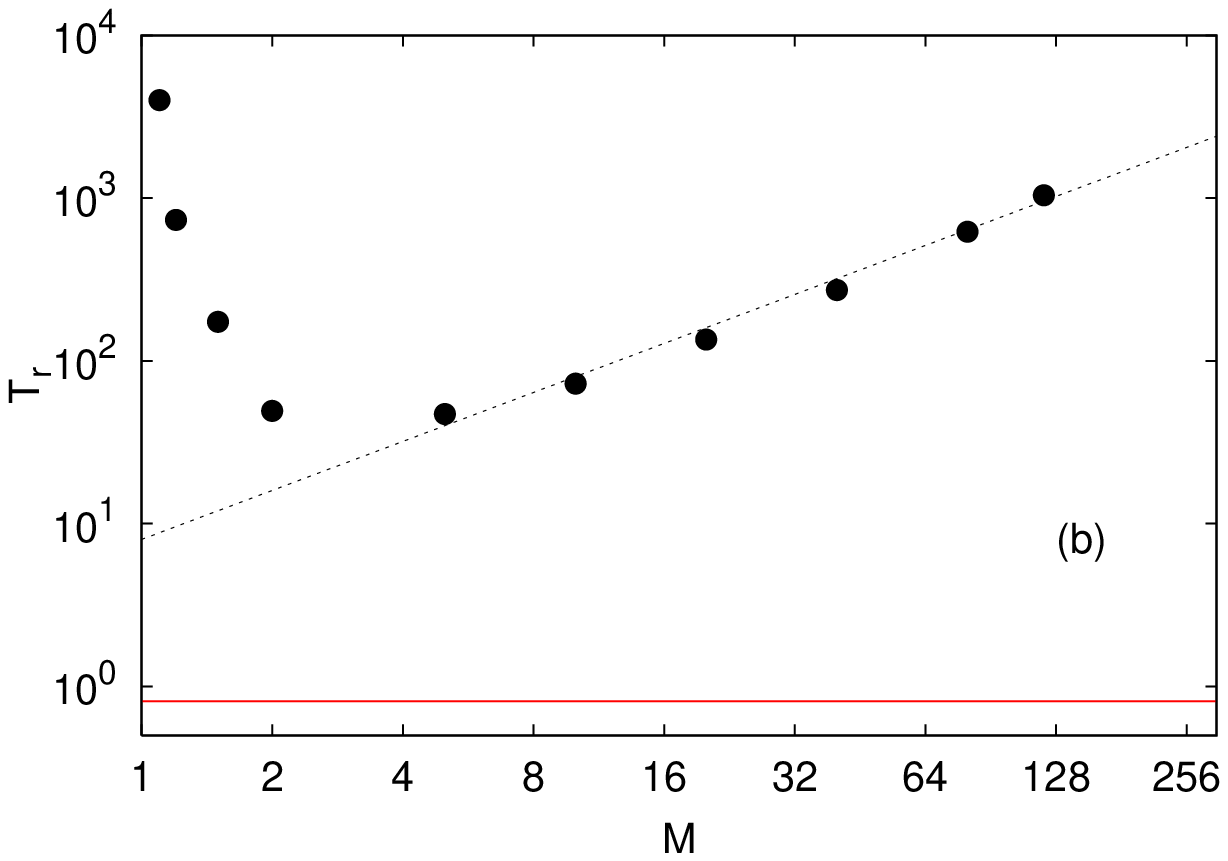}
\caption{ (a) $\langle\mathcal{K}(t)\rangle_e$ vs time for HD with
  different impurity mass values including $M=m$ (i.e. absence of the
  colloid). All curves collapse confirming that in the case of disks
  the relaxation time is independent of the presence of the colloid.
  The black curve represents the fit (\ref{eq:fit}). (b) Relaxation
  time $\mathcal{T}_r$ vs $M$ needed for HPS to relax from a uniform to the
  Gaussian equilibrium distribution at $T=50$. Other parameters are
  fixed as in Fig.~\protect\ref{fig:diffsquares}. The dotted straight
  line suggests a compatibility with a linear behavior at large mass
  ratios. The red horizontal line indicates $\mathcal{T}_r$ for the HD, which is
  put here for comparison and to show the independence of relaxation
  process on the presence of the colloid as already evidenced in (a).
}\label{fig:relax}
\end{figure}

In our simulations, we prepare the HD or HPS gas in a state
with a spatially homogeneous distribution of particles and
flat velocity distribution, i.e. $P(v_x,v_y) = g(v_x)g(v_y)$ with
$g(v)= 1/(2 v_0)$ for $|v| < v_0$ and zero elsewhere. The value of
$v_0$ is fixed by imposing the temperature $T$ of the system, i.e.
$v_0^2/6 = \int dv g(v) v^2 = T/m$.  The colloidal particle is
initialized with random velocity extracted from the same distribution.
The system is then let evolve under the event driven dynamics, and the
velocity pdf monitored by computing
$$ \mathcal{K}(t) = \frac{\overline{v_x(t)^4}}
{3\overline{v_x(t)^2}^2} = \frac{\overline{v_y(t)^4}}
{3\overline{v_y(t)^2}^2}.$$ The symbol $\overline{[\dots]}$ indicates
the average at a given time $t$ over the  gas particles,
i.e. $\overline{v_x^2} = 1/N \sum_{i=1}^{N} v_{xi}^2$.  At $t=0$,
$\mathcal{K}=K_0=3/5$ while for $t\to \infty$ the Gaussian result
$\mathcal{K}=K_\infty=1$ should hold, being the system relaxed.  To
have a smooth behavior we average $\mathcal{K}(t)$ over many
independent runs. We thus obtain $\langle
\mathcal{K}(t)\rangle_e$ which is shown in Fig.~\ref{fig:relax}a for
HD and is well described by the fitting function
\begin{equation}
\langle \mathcal{K}(t) \rangle_e =  K_\infty  + 
(K_0-K_\infty) \mbox{e}^{-t/\mathcal{T}_r}  
\label{eq:fit}
\end{equation}
where the fitting parameter $\mathcal{T}_r$ provides an estimate of the
relaxation time to equilibrium of the system. The brackets $\langle
[\dots]\rangle_e$ denote averages over the realizations.

The perfect collapse of the curves $\langle \mathcal{K}(t) \rangle_e$
for the HD system (Fig.~\ref{fig:relax}a) obtained for four values of
the ratio $M$ including the case $M=m=1$ (i.e. without the impurity),
clearly indicates that the relaxation process of the disks is
independent of the colloid.  Unlike HD, for HPS the mass $M$ of the
impurity is crucial in determining the relaxation. This is shown in
Fig.~\ref{fig:relax}b, where we report the behaviour of $\mathcal{T}_r$, fitted
by using (\ref{eq:fit}), as a function of $M$. As one can see, $\mathcal{T}_r$
diverges for $M \to m$ (identical squares) and $M\to \infty$
(immobile impurity which again does not allow for the exchange of
energy, leading to the impossibility of relaxation). Notice that
asymptotically $\mathcal{T}_r$ seems to grow linearly with $M$.

This results, although obvious when considering the different nature
of the collisional processes occurring in the two systems, confirm
that the relaxation to the equilibrium of HPS is much slower than that
for HD and crucially depends on the impurity characteristics.

\section{Final remarks}
\label{sec:discuss}

From the results presented in this paper we obtain good indications
that a non-chaotic system with many degrees of freedom, consisting of
a gas of hard squares, provides a suitable  model for
Brownian motion which is equivalent, at least at a simulation level,
to the corresponding chaotic model, where squares are replaced by
disks.

A deep understanding the role of chaos for Brownian motion and for
transport properties is a very important issue deserving further
comments.  With reference to previous works, we now discuss this issue
by  stressing how the presence/absence of ergodicity, mixing and
chaos in the microscopic dynamics (may) influence the statistical
mechanics of macroscopic systems.

Let us start with a few remarks on equilibrium statistical mechanics,
where the problem of connecting (micro)dynamics and (macro)statistical
features is at the origin of Boltzmann's ergodic hypothesis. First, it
is worth noticing that ergodicity, in its strict mathematical
formulation, is an extremely demanding property. Second, in spite of
its theoretical importance, it is not completely satisfactory from a
physical point of view as it involves global asymptotic limits, rarely
encountered in practice. However, for the foundation of statistical
mechanics, a widespread consensus exists on the key role played by the
huge number of degrees of freedom involved in a macroscopic system
rather than ergodicity.  This point of view received mathematical
support from the works by Khinchin~\cite{Khinchinbook}, Mazur and van
der Linden \cite{Mazur}, and others. They proved that statistical
mechanics works independently of ergodicity thanks to the existence of
meaningful physical observables (the so-called sum functions) which
are nearly constant on the energy surface, apart from regions of
vanishing measure. Support to this picture
comes from the results of Frisch and coworkers
\cite{Frisch1,Frisch2,Frisch3,Frisch4}, who found robust statistical
phenomena in a trivially non-ergodic system. Nevertheless, Khinchin's
statements cannot be the ultimate grounding of statistical mechanics
because not all physically important observables belong to the
class of the sum functions.

Since chaos grants the validity of some ``statistical laws'' even in
few degrees of freedom systems, one could be tempted to invoke it as
the sufficient ingredient to build a robust statistical mechanical
approach grounded on Hamiltonian systems. However, for this to be
true, chaos in the microscopic dynamics should be enough
``strong''. Indeed, results\footnote{In chaotic high dimensional
systems one may have a sort of ``localized chaos'' without a globally
irregular dynamics. This phenomenon is somehow the high dimensional
analogous to the presence of chaos in bounded regions with the absence
of large scale diffusion observed in low dimensional symplectic
systems in situations below the resonance overlap~\cite{Chirikov}.}
of extended simulations in high dimensional systems
\cite{Livi,Benettin,Casartelli} have shown that chaos may be not
enough to ensure the validity of the equilibrium statistical
mechanics.

Let us now discuss this issue in the non-equilibrium statistical
mechanics context, where the analogous of ergodicity is the mixing
condition. As before, this condition is very demanding, being related
to the $\Gamma$-space (the set of positions and momenta of system
particles); while the study of macroscopic systems usually focuses on
physical observable involving some projection procedures, amounting to
neglect (or average) the effects of a large number of degrees of
freedom in favor of few relevant variables. For instance, in the case of
elementary transport properties the observable that matters refer to
single particle properties: mean square particle displacement,
correlation function and the response function either of the colloidal
particle or the single gas particle. Therefore, one can wonder about
the microscopic conditions ensuring a ``good'' statistical behavior
for the above quantities.

We can start by considering few degrees of freedom systems, where also
simple deterministic chaotic models may exhibit transport properties
similar to those of more realistic systems.  Paradigmatic examples are
chaotic billiards and the Lorentz gas, where particle trajectories are
chaotic as a consequence of the convexity of the obstacles.  Numerical
and theoretical works have shown that, in these systems (under
appropriate hypothesis such as hyperbolicity etc.) the transport
coefficients can be quantitatively related to chaos
indicators~\cite{gaspnico,viscardy}. This would suggest that chaos is
tightly related to transport.  However, several examples of
non-chaotic deterministic systems, such as a bouncing particle in a
two-dimensional billiard with polygonal randomly distributed
obstacles, possess robust transport properties
\cite{dettmann,dettmann2,cecconi,casati,bambihu}. For these
non-chaotic models, it has been proposed that a sort of non-linear
instability mechanism is required to observe
diffusion~\cite{cecconi}. The existence of non-chaotic models able to
display diffusion poses some doubts on the possibility to
make  strong statements on the role of chaos for
transport. It is however important to stress that in all these
models the particles do not interact, and therefore, at least from a
statistical mechanics point of view, they are rather artificial.

More interesting it is thus to consider many degrees of freedom
systems, such a those investigated in this work. In this case the
existence of quantitative relationships among chaos indicators and
transport coefficients is, to the best we know, less clear (see for
example \cite{barnett,torcini}). Nevertheless, chaos has been proved
to be relevant in the establishment of some non-equilibrium properties
\cite{RONDOREFS}.  Moreover, it is fair to say that diffusion of
particles is rather common in chaotic many body particle systems.  On
the other hand, the non-chaotic model here investigated together with
the previous results by Frisch and coworkers
\cite{Frisch1,Frisch2,Frisch3} indicate that transport properties for
both impurity and gas particles agree with the prediction of kinetic
theory and are indistinguishable from those of the (mixing) hard disk
model.  Therefore chaos, at least in the sense of positive Lyapunov
exponents, cannot be invoked to explain the observed statistical
behaviors. Nevertheless, as for low dimensional models \cite{cecconi},
also in HPS a non infinitesimal mechanism of instability can be
induced by the presence of singular corners of the squares, and these
likely play a role for the diffusive behaviour.

We interpret these findings in the framework developed by Khinchin:
the ``good transport'' properties observed in the non-mixing system
result from the large number of particles and not from chaos.  This is
well evident for the correlation and response functions for the hard
squares when, e.g., all particle are identical and the collision
dynamics reduces to a mere relabelling.  In such a case the
exponential relaxation of $C_v(t)$ and $R_v(t)$ is just a
probabilistic consequence of the exponential distribution of the time
interval between two consecutive collisions.  We stress that here
diffusive properties are the outcome of the action of many degrees of
freedom, and not of non-linear instability mechanisms as in low
dimensional chaotic and non-chaotic models.  Of course, as discussed
in Sect.~\ref{sec:relaxfar}, chaos may favour the equilibration of the
system.

As a last remark we note that the Gallavotti-Cohen \cite{galla}
  fluctuation theorem seems to apply to non-chaotic models, at least
  in finite time intervals, as shown by Benettin et al. \cite{bene}
  who investigated a non-equilibrium version of the Ehrenfest
  wind-tree model, which is non-chaotic.  In such a system, although
  the maximum Lyapunov exponent is zero, the presence of long
  irregular transients, introduces an ``effective randomness''.

In conclusion, we think that it is very difficult to decipher the
signature of chaos in transport phenomena observed in many particles
systems because, as shown in this paper, it can be overwhelmed by the
emergence of an ``effective dynamical randonmess'' due to the
combination of: a) coarse-graining procedure, b) finite scale
instability and c) presence of a huge number of degrees of freedom.
The characterization of this ``effective randomness'' requires the
renounce to asymptotic limits (arbitrarily long time and arbitrary
resolution) in favour of a finite time and/or finite resolution
analysis \cite{ford2,dettmann,wang,boffy}.

\ack{We thank O. Jepps and L. Rondoni for useful discussions,
  suggestions and their contribution in early stage of this work. We
  acknowledge U. Marini Bettolo Marconi for discussions and pointing
  out some useful references. We thank M. Falcioni, S. Pigolotti and
  A. Puglisi for careful reading the manuscript and useful remarks.
  This work has been partially supported by the PRIN2005 ``Statistical
  mechanics of complex systems'' by MIUR. }


\appendix 
\section{Computation of the diffusion coefficient}
In this Appendix, we detail the computation of the diffusion
coefficient for the colloidal particle in an uniform and rarefied HD
and HPS gas, following elementary kinetic theory.  The basic idea is to
estimate the average drag force exerted by the gas particles which
collide with the impurity, by calculating the average exchanged
momentum in the collisions.

\begin{figure}[t!]
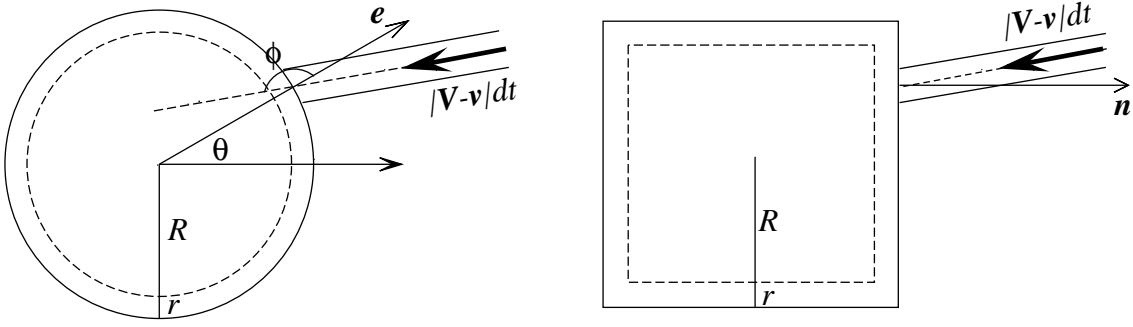

\includegraphics[width=.5\textwidth,keepaspectratio,clip=true]{Fig8a}
\includegraphics[width=.5\textwidth,keepaspectratio,clip=true]{Fig8b}
\caption{Cartoon of geometric construction used to compute the collision 
rate between an HD and HPS colloidal particle characterized by 
$(M,R,{\bm V})$, and gas particles $(m,r,{\bm v})$. 
The equivalent problem considers the colloid at rest, 
but with increased size $R+r$, and 
pointlike gas particles moving at the relative 
velocity ${\bm v} - {\bm V}$. The collision rate amount counting 
the number of gas particles contained in the collisional cylinder.}
\label{fig:cartoon}
\end{figure}

Consider a rarefied HD gas at equilibrium, and focus on the collision
of the colloidal disk characterized by its mass $M$, radius $R$ and
precollisional velocity $\bm V$ with the gas particles which are
characterized by $m,r,\bm v$, respectively.  According to
Eq.~(\ref{eq:hdcrule}), the implulse transferred in the collision is
$$
M {\bm\Delta}{\bm V} = M ({\bm V}' - {\bm V}) = \frac{2mM}{m+M} {\bm g} \,,
$$ ${\bm g} = {\bm v} - {\bm V}$ being the precollisional relative
velocity.  The rate of such collisions can be obtained by
considering the equivalent problem of a colloid, at rest, with radius
$r+R$, and hit by a flux of pointlike particles moving at relative
velocity ${\bm g}$.  The rate is then determined by counting the
number of pointlike particles hitting the unit surface per unit time
for a given orientation ${\bm e}$.  This number corresponds to the
particles contained in the collisional cylinder of infinitesimal base
$(R+r)d \theta$ and height $\rho |{\bm g} \cdot {\bm e}| \Theta(-{\bm
e}\cdot {\bm g}) \delta t$, as shown in Fig.~\ref{fig:cartoon}.  The
unitary step function $\Theta(s)$ selects the condition, $\bm
g\cdot\bm e<0$, to have a collision.

Accordingly, the mean impulsive force in the normal direction ${\bm e}
= \{\cos(\theta),\sin(\theta)\}$ selected by the $\theta$-angle that
${\bm e}$ forms with vector ${\bm V}$ (taken as $x$-axis direction)
\begin{equation}
\bigg \langle M \frac{\delta\bm V_n}{\delta t} \bigg\rangle =
\frac{2mM}{m+M} \rho (R+r)\int_{0}^{2\pi}\!\!\! d\theta \int \!\!d{\bm v} P({\bm
v})\Theta(-{\bm e}\cdot {\bm g}) |{\bm g} \cdot {\bm e}| ({\bm
g}\cdot{\bm e}){\bm e}\,, \quad
\label{eq:integr}
\end{equation}
the average is meant over the equilibrium distribution of the gas
velocities $P(\bm v) = m/(2\pi T)\exp[-m |{\bm v}|^2/(2T)]$.  It is
convenient to make the change of variable $(v_x,v_y) \to (g_x,g_y)$
and then perform the approximation $P({\bm v}) \simeq P({\bm g})[1 -
m/T {\bm g} \cdot {\bm V}]$, justified in the limit $M \gg m$.
After this manipulation, the integral (\ref{eq:integr}) becomes
$$
\bigg \langle 
M \frac{{\delta}{\bm V}_n}{\delta t} 
\bigg \rangle 
\simeq 
-\frac{4m^2M}{T(m+M)}  \rho (R+r)
\int_0^{2\pi}\!\! d\theta \!\int\!\! d{\bm g}P({\bm g}) \Theta(-{\bm e}\cdot{\bm g})
({\bm g}\cdot{\bm e})^2 ({\bm g}\cdot{\bm V})\,{\bm e}\;.\quad
$$ This expression, recast in the form $M\dot{{\bm V}} = -\gamma M
{\bm V}$, allows to explicitate the friction coefficient $\gamma$.
Passing to polar coordinates, $g_x=g\cos(\alpha),g_y=g\sin(\alpha)$,
simplifies the integral ($\alpha$ being the angle between ${\bm g}$
and ${\bm V}$, whose direction coincides with $x$-axis).  Indicating by
$\phi$ the angle between ${\bm e}$ and ${\bm g}$, we end up with
$$
\int_{0}^{2\pi} d\theta \int_{0}^{\infty} dg P(g) g^4 \int_{0}^{2\pi} d\alpha
|\cos(\phi)|cos(\phi)\Theta[-\cos(\phi)] V\cos(\alpha)
\,\cos(\theta)\;.      
$$ where the angles $\alpha,\theta,\phi$ are related by $\theta-\alpha
= \pi -\phi$. The integration over $g$ yields
$3/(2\pi)\sqrt{\pi/2}(m/T)^{-3/2}$, and that one on the angles gives
the value $4\pi/3$. We then obtain the friction coefficient 
\begin{equation}
\gamma= 2 \sqrt{2\pi}\, \frac{\rho R \sqrt{ mT}}{M} \left(\frac{1+{r}/{R}}{1+{m}/{M}}\right)\,.
\label{eq:gammaHD}
\end{equation}
where the last factor takes into account finite size and mass corrections.
By using Eq.~(\ref{II.7})  the diffusion constant is obtained as
\begin{equation}
\label{eq:dcHD}
D_c = \frac{T}{M\gamma} = 
\frac{1}{2\sqrt{2\pi}}\frac{1}{\rho R}\sqrt{\frac{T}{m}}
\left(\frac{1+m/M}{1+r/R}\right)
\end{equation} 

The same computation can be repeated for HPS, the only difference lying
in the geometry of the problem (Fig.~\ref{fig:cartoon}).  Thanks to
the symmetry, it is convenient to decompose the average impulse
transferred from the gas to the colloidal particle in the x and y
components (along the directions $(1,0)$, or $(0,1)$). For example, in
the x-direction we have
$$ \bigg \langle M \frac{{\delta V}_x} {\delta t} \bigg\rangle \simeq
- \frac{4m^2 M}{T(m+M)}  \rho (R+r) V_x \sum_{n_x=\pm 1} \int \!\!dg_x
P(g_x) \Theta(-g_x n_x) (g_x n_x)^2 |g_x|
$$ Notice that the sum over $n_x=\pm 1$ stems from the two identical
contributions to the exchanged momentum given by the left and right
collisions occurring on the opposite sides of the square.  The
integral can be solved in Cartesian coordinates, and considering that
the constraint imposed by $\Theta[-g_x n_x]$ amounts to a factor
$2$. We thus obtain
$$ \bigg \langle M \frac{{\delta V}_x} {\delta t} \bigg\rangle
=-\frac{8m^2M}{T(m+M)}\rho (R+r) V_x \int_{0}^{\infty} dg_x P(g_x)
g_x^3\;.
$$
The integration gives the value $\sqrt{2/\pi}(m/T)^{3/2}$ 
yielding the final results for the friction coefficient
\begin{equation}
\gamma= 8\sqrt{\frac{2}{\pi}} \frac{\sqrt{m}\rho R \sqrt{T}}{M} \left(\frac{1+{r}/{R}}{1+{m}/{M}}\right)\,,
\label{eq:gammaHS}
\end{equation}
and thus the diffusion coefficient reads
\begin{equation}
\label{eq:dcHS}
D_c = \frac{\sqrt{\pi}}{8\sqrt{2}}\frac{1}{\rho R}\sqrt{\frac{T}{m}}  \left(\frac{1+m/M}{1+r/R}\right)\,.
\end{equation}



\section*{References}
\begin{thebibliography}{250}
\bibitem{Einstein} Einstein A, 
\textit{Ueber die von der
  molekularkinetischen Theorie der Waerme geforderte Bewegung von in
  ruhenden Fluessigkeiten suspendierten Teilchen}, 
1905  \textit{Ann. Phys.} \textbf{17} 549

\bibitem{Smolu} Smoluchowski M,
\textit{Zur kinetischen Theorie der Brownschen Molekularbewegung und der Suspensionen}, 
1906  \textit{Ann. Phys.} \textbf{21} 756

\bibitem{DetDiff} Geisel T and Nierwetberg J, 
\textit{Onset of  Diffusion and Universal Scaling in Chaotic Systems}, 
1982  \textit{Phys. Rev. Lett.} \textbf{48} 7

\bibitem{DetDiff2} Grossmann S and Fujisaka H, 
\textit{Diffusion in  discrete nonlinear dynamical systems},
1982 \textit{Phys. Rev. A}  \textbf{26} 1779


\bibitem{Klages} Klages R and Dorfman J R, 
\textit{Simple maps with  fractal diffusion coefficients}, 
1995 \textit{Phys. Rev. Lett.}   \textbf{74} 387


\bibitem{gaspnico} Gaspard P and Nicolis G, 
\textit{Transport  properties, Lyapunov exponents, and entropy per unit time}, 
1990  \textit{Phys. Rev. Lett.} \textbf{65} 1693

\bibitem{viscardy} Viscardy S and Gaspard P, 
\textit{Viscosity in the escape-rate formalism},
2003 \textit{Phys. Rev. E} \textbf{68} 041205

\bibitem{GaspDor} Gaspard P and Dorfman J R, 
\textit{Chaotic scattering theory, thermodynamic formalism, and transport coefficients}, 
1995 \textit{Phys. Rev. E} \textbf{52} 3525


\bibitem{dorfman} Dorfman J R 
\textit{An Introduction to Chaos in Nonequilibrium Statistical Mechanics}, 
1999 (Cambridge: Cambridge University Press)


\bibitem{gaspard} Gaspard P, 
{\it Chaos, scattering and statistical  mechanics}, 
1998 (Cambridge: Cambridge University Press)


\bibitem{ford2} Vega J L, Uzer T, Borondo F and Ford J,
\textit{Deterministic diffusion in almost integrable systems}
1996 \textit{Chaos} \textbf{6} 519


\bibitem{dettmann} Dettmann C P and   Cohen E D G,
\textit{Microscopic chaos and diffusion},
2000 \textit{J. Stat. Phys.} \textbf{101} 775

\bibitem{dettmann2} Dettmann C P and   Cohen E D G,
\textit{Note on chaos and diffusion},
2001  \textit{J. Stat. Phys.}  \textbf{103} 589 

\bibitem{cecconi}Cecconi F, del-Castillo-Negrete D,  Falcioni M and Vulpiani A, 
\textit{The origin of diffusion: the case of non-chaotic systems}, 
2003 \textit{Physica D} \textbf{180} 129; 
Cecconi F,  Cencini M, Falcioni M and Vulpiani A,
\textit{Brownian motion and diffusion: from stochastic processes to chaos and beyond}
2005  \textit{Chaos} \textbf{15} 026102


\bibitem{LLP} Lepri S, Livi R and Politi A,
\textit{Thermal conduction in classical low-dimensional lattices},
2003 \textit{Phys. Rep.} \textbf{377} 1


\bibitem{casati} Alonso D,  Artuso R,  Casati G and  Guarneri I,
\textit{Heat conductivity and dynamical instability},
1999 \textit{Phys. Rev. Lett.} \textbf{82} 1859

\bibitem{casati2}  Li B, Casati G and  Wang J,
\textit{Heat conductivity in linear mixing systems},
2003 \textit{Phys. Rev. E} \textbf{67} 021204

\bibitem{bambihu} Li B, Wang L and Hu B,
\textit{Finite Thermal Conductivity in 1D Models Having Zero Lyapunov Exponents},
2002 \textit{Phys. Rev. Lett.} \textbf{88} 223901 


\bibitem{grassberger}  Grassberger P,  Nadler W and  Yang L,
\textit{Heat conduction and entropy production in a one-dimensional hard-particle gas}, 
2002 \textit{Phys. Rev. Lett.} \textbf{89} 180601 

\bibitem{fuzzy} Cecconi F, Livi R and Politi A, \textit{Fuzzy
transition region in a 1D coupled-stable-map lattice} 1998
\textit{Phys. Rev. E} \textbf{57} 2703

\bibitem{Gaspard1994} Gaspard P., \textit{Comment on dynamical randomness in
quantum systems} 1994 \textit{Prog. Theor. Phys. Suppl.} \textbf{116} 369

\bibitem{prigogine}  Prigogine I and  Stengers I,
\textit{Entre le temps et l'\'eternit\'e},
1979 (Paris: Fayards);
Prigogine I,
\textit{Laws of Nature, Probability and Time Symmetry Breaking},
1999 \textit{Physica A} \textbf{263} 528

\bibitem{Khinchinbook} Khinchin A I,
 \textit{Mathematical Foundations of Statistical Mechanics},
1949  (New York: Dover Publications Inc.)

\bibitem{Mazur}Mazur P and van der Linden J,
\textit{Asymptotic form of the structure function for real systems},
1963 \textit{J. Math. Phys.}  \textbf{4} 271

\bibitem{Bricmont}  Bricmont J,
\textit{Science of chaos or chaos in science?},
1996 \textit{Ann. New York Ac. of Sciences} \textbf{775} 131

\bibitem{Holley} Holley R,
\textit{Motion of a heavy particle in an infinite one dimensional gas of hard spheres},
1971 \textit{Zeit. Wahrsch. Verw. Geb.} \textbf{17} 181

\bibitem{Lebovitz} D\"urr D, Goldstein S and  Lebowitz J L,
\textit{A mechanical model of Brownian motion},
1981 \textit{Comm. Math. Phys.}  \textbf{78} 507

\bibitem{Frisch1} Szu H H, Bdzil J, Carlier C, and Frisch H L,
\textit{Molecular-dynamics verification of a final velocity distribution of a 
nonergodic system of hard parallel squares},
1974 \textit{Phys. Rev. A} \textbf{9} 1359

\bibitem{Frisch2} Frisch H L, Roth J, Krawchuk B D, and Sofinski P,
\textit{Molecular dynamics of nonergodic hard parallel squares with a 
Maxwellian velocity distribution}  
1980 \textit{Phys. Rev. A} \textbf{22} 740
 

\bibitem{Frisch3} Carlier C and Frisch H L,
\textit{Molecular Dynamics of Hard Parallel Squares}
1972 \textit{Phys. Rev. A} \textbf{6} 1153;
\textit{Molecular-Dynamics Study of Clustering in Hard Parallel Squares}
1973 \textit{Phys. Rev. A} \textbf{7} 348

\bibitem{Frisch4}  Rudd W G, and Frisch H L,
\textit{The equation of state of parallel hard squares},
1971 \textit{J. Comp.Phys} \textbf{7} 394

\bibitem{Kubo}  Kubo R,
\textit{Brownian motion and nonequilibrium statistical mechanics},
1986 \textit{Science} \textbf{233} 330 

\bibitem{kubobook}  Kubo R, Toda M and Hashitsume N,
\textit{Statistical Physics}, vol 2.
1985 (Berlin: Springer-Verlag)


\bibitem{Grassia} Grassia P, 
\textit{Dissipation, fluctuations and conservation laws},
2001 \textit{Am. J. Phys.} \textbf{69} 113

\bibitem{AT93} Allen M P and Tildesley T J, 
\textit{Computer  simulation of Liquids},
1993 (Oxford: Clarendon Press)

\bibitem{REVIEWGENERALE} Garcia-Rojo R, Luding S, and Brey J J,
\textit{Transport coefficients for dense hard-disk systems},
2006 \textit{ Phys. Rev. E} \textbf{74} 061305 

\bibitem{LorentzGas}  Lorentz H A,
\textit{The motion of electrons in metallic bodies},
1905 \textit{Proc. Amst. Acad.} \textbf{7} 438, 585, 684

\bibitem{Posch} van Beijeren H, Dorfman J R, Cohen E G D, Posch H A, and Dellago C, 
\textit{Lyapunov Exponents from Kinetic Theory for a Dilute, Field-Driven Lorentz Gas},
1996 \textit{Phys. Rev. Lett.} \textbf{77} 1974

\bibitem{Alder} Alder B J and Wainwright T E, 
\textit{Decay of the Velocity Autocorrelation Function},
1967 \textit{Phys. Rev. Lett.} \textbf{18}  988 

\bibitem{Cohen_et_al} Dorfman J R and Cohen E G D,
\textit{Velocity Correlation Functions in Two and Three Dimensions}
1970 \textit{Phys. Rev. Lett.} \textbf{25} 1257

\bibitem{longtimes} Perondi L F and  Binder P M, 
\textit{Mean-squared displacement of a hard-core tracer in a periodic lattice},
1993 \textit{Phys. Rev. B} \textbf{48} 4136

\bibitem{LeboCherno} Chernov N and  Lebowitz J L,
\textit{Dynamics of a Massive Piston in an Ideal Gas: Oscillatory Motion and Approach to Equilibrium}, 
2002 \textit{J. Stat. Phys.} \textbf{109} 507


\bibitem{A93} Ackland G J, 
\textit{Equipartition and ergodicity in closed one-dimensional systems of hard spheres with different masses}, 
1993 \textit{Phys. Rev. E} \textbf{47} 3268

\bibitem{SBJ06} Shirts R B, Burt S R and Johnson A M, 
\textit{Periodic boundary condition induced breakdown of the equipartition principle and other kinetic effects of finite sample size in classical hard-sphere molecular dynamics simulation}, 
2006  \textit{J. Chem. Phys.} \textbf{125} 164102

\bibitem{Chirikov} Chirikov B V,
\textit{Universal instability of many-dimensional oscillator systems},
1979 \textit{Phys. Rep.} \textbf{52} 263 

\bibitem{Livi} Livi R,  Pettini M, Ruffo S and  Vulpiani A,
\textit{Chaotic behaviour in nonlinear Hamiltonian systems and equilibrium statistical mechanics},
1987 \textit{J. Stat. Phys.} \textbf{48} 539 

\bibitem{Benettin} Benettin G, Galgani L, Giorgilli A,
\textit{Boltzmann ultraviolet cutoff and Nekhoroshev theorem on Arnold  diffusion}  
1984 \textit{Nature} \textbf{311} 444 

\bibitem{Casartelli} Alabiso C, Casartelli M, Marenzoni P,
\textit{Thermodynamic limit beyond the stochasticity threshold in nonlinear chains},
1993 \textit{Phys. Lett. A} \textbf{183} 305 

\bibitem{barnett} Barnett D M, Tajima T, Nishihara K, Ueshima Y, and
Furukawa H, \textit{Lyapunov Exponent of a Many Body System and Its
Transport Coefficients}, 1996 \textit{Phys. Rev. Lett.} \textbf{76}
1812

\bibitem{torcini} Torcini A,  Dellago C and  Posch H A,
\textit{comment on ``Lyapunov exponent of a many body system and its transport coefficients}
1999 \textit{Phys. Rev. Lett} \textbf{83}  2676

\bibitem{RONDOREFS} Evans D J, Cohen E G D and Morriss G P, 
\textit{Viscosity of a simple fluid from its maximal Lyapunov exponents}, 
1990 \textit{Phys. Rev. A} \textbf{42} 5990;
Sarman S,  Evans  D J and    Morriss G P,
\textit{Conjugate-pairing rule and thermal-transport coefficients},
1992 \textit{Phys. Rev. A} \textbf{45} 2233

\bibitem{galla} Gallavotti G and Cohen E G D, \textit{Dynamical
  Ensembles in Nonequilibrium Statistical Mechanics}, 1995
  \textit{Phys. Rev. Lett.} \textbf{74} 2694

\bibitem{bene} Lepri S, Rondoni L and Benettin G, \textit{The
  Gallavotti–Cohen Fluctuation Theorem for a Nonchaotic Model}, 2000
  \textit{J. Stat. Phys.} \textbf{99} 857

\bibitem{wang} Gaspard P and Wang X J, 
\textit{Noise, chaos and $(\epsilon,\tau)$-entropy per unit time}, 1993 
\textit{Phys. Reports} \textbf{235} 291

\bibitem{boffy}
 Boffetta G, Cencini M, Falcioni M and Vulpiani A,
\textit{Predictability: a way to characterize complexity}, 2002 
\textit{Phys. Reports} \textbf{356} 367 


\end{thebibliography}
\end{document}